\documentclass[pra,superscriptaddress]{revtex4}

\usepackage{graphicx}
\usepackage{amsmath}
\usepackage{amssymb}
\usepackage{hyperref}\hypersetup{colorlinks=true,linkcolor=blue,citecolor=blue,pdfstartview=FitH}

\DeclareMathOperator{\ci}{ci}
\DeclareMathOperator{\cosech}{cosech}

\DeclareMathOperator{\sign}{sign}
\DeclareMathOperator{\Tr}{Tr}

\DeclareMathOperator{\Real}{Re}\renewcommand{\Re}{\Real}

\begin{document}
\title{Quantum quench dynamics of the sine-Gordon model in some solvable limits}
\author{A.~Iucci}
\affiliation{Instituto de F\'{\i}sica la Plata (IFLP) - CONICET and Departamento de F\'{\i}sica,\\
Universidad Nacional de La Plata, CC 67, 1900 La Plata, Argentina}
%
%
\affiliation{Donostia International Physics Center (DIPC), Manuel de Lardiz\'abal 4, 20018 San Sebasti\'an, Spain}
\author{M.~A.~Cazalilla}
\affiliation{Centro de F\'{\i}sica de Materiales (CFM). Centro Mixto CSIC-UPV/EHU. Edificio Korta, Avenida de Tolosa
72, 20018 San Sebasti\'an,  Spain}
\affiliation{Donostia International Physics Center (DIPC), Manuel de Lardiz\'abal 4, 20018 San Sebasti\'an, Spain}
\begin{abstract}
In connection with the the thermalization problem in isolated quantum systems, we investigate the dynamics following a quantum quench of the sine-Gordon model in the Luther-Emery and the semiclassical limits. We consider the quench from the gapped to the gapless phase  as well as reversed one. By obtaining analytic expressions for the one and two-point correlation functions of the order parameter operator at zero-temperature, the manifestations of integrability in the absence of thermalization in the sine-Gordon model are studied. It is thus shown that correlations in the long time regime after the quench  are well described by a generalized Gibbs ensemble. We also consider the case where the system is initially in contact with a reservoir at finite temperature.  The possible relevance of our results to current and future experiments with ultracold atomic systems is also critically considered.
\end{abstract}
\maketitle
\section{Introduction}

Recent experiments in the field of ultracold atomic gases have spurred much
interest in understanding the thermalization dynamics of isolated quantum systems.%
\cite{altman02_quench,%
sengupta04_quench_QCP,%
yuzbashyan05_BCS_quench,%
kollath05_density_waves,%
altman05_projection_feshbach_molecules,%
kinoshita06_non_thermalization,%
ruschhaupt06_quench_momentum_interference,%
barankov06_dynamical_projection,%
calabrese06_quench_CFT,%
rigol06_hcbosons_dynamics,%
cazalilla06_quench_LL,%
manmana07_quench_spinless_fermions_nnn,%
kollath07_quench_BH,%
rigol07_generalized_gibbs_hcbosons,%
calabrese07_quench_CFT_long,%
gritsev07_spectroscopy_quench,%
degrandi08_adiabatic_quench,%
eckstein08_generalized_gibbs_FK,%
kollar08_generalized_gibbs_hubbard,%
moeckel08_quench_hubbard,%
rigol08_mechanism_thermalization,%
sen08_nonlinear_quench_QCP,%
roux08_time_dependent_lanczos,%
reimann08_foundations_stat_mech,%
faribault09_fermion_pairing_model,%
iucci09_quench_LL,%
sotiriadis09_quench_finite_T,%
iucci09_quench_preprint}  So far, the latter were considered mere
idealizations of real systems, as most of the many particle systems
of interest to quantum statistical mechanics, such as solids and quantum
fluids, are strongly coupled to their environments. However, the creation of large ensembles of
ultracold atoms  with highly controllable properties, which remain fully quantum
coherent for relatively long times (compared to the typical duration of an experiment),
has completely changed this perception. This has also raised concerns about the
mechanisms of thermalization  in these systems, and even in some recent experiment~\cite{kinoshita06_non_thermalization}
lack of thermalization has been observed.

The problem of  thermalization in isolated  quantum systems can be also posed as the study of the dynamics of the system following a  \emph{quantum quench}. That is to say, the study of the response of the system to a change of a control parameter of the Hamiltonian over a time-scale much shorter than any other relevant time scale of the system, so that the sudden approximation can be applied. Therefore, it is assumed that for $t<0$ the Hamiltonian is $H_\text{i}$ and the system is in a given eigenstate  of it, $\vert\Phi_\text{i}\rangle$. At $t=0$ the Hamiltonian is changed to $H_\text{f}$ and thus for $t>0$ the system evolves unitarily in isolation according to the dynamics  dictated by $H_\text{f}$. The quantum quench can be also used to describe the evolution of a system that has been prepared in a given state that is \emph{not}   an eigenstate of the Hamiltonian. Thus, the question  naturally arises is whether a set of sufficiently interesting observables  of the system reach some form of stationary state that can be described by a standard Gibbs  ensemble. In such a case, we would speak of  thermalization to a standard statistical ensemble (microcanonical, canonical, or grand-canonical). However,  this may not be the case as it turns out that for several integrable
models~\cite{rigol07_generalized_gibbs_hcbosons,cazalilla06_quench_LL,calabrese07_quench_CFT_long,%
kollar08_generalized_gibbs_hubbard,iucci09_quench_LL} the long-time behavior of certain observables can instead obtained from  a generalized Gibbs ensemble characterized by a different temperature for each eigenmode of the Hamiltonian $H_\text{f}$~\cite{rigol07_generalized_gibbs_hcbosons,cazalilla06_quench_LL,iucci09_quench_LL,sotiriadis09_quench_finite_T,%
iucci09_quench_preprint}. The non-standard or
generalized Gibbs ensembles follow from maximizing the von Neumann entropy  with the constraints imposed by the set integrals of motion larger than the total energy and particle number.  As to the situation concerning
non-integrable systems, the issue of the thermalization dynamics is still not fully understood although some
recent results~\cite{rigol08_mechanism_thermalization,reimann08_foundations_stat_mech,moeckel08_quench_hubbard}
indicate that thermalization to a standard ensemble should eventually occur at sufficiently long times. However,
the actual thermalization dynamics and how it depends to how close the system is
to integrability still remain very poorly understood.  Indeed, as far as one dimensional systems are concerned (for which the strong kinematic constrains usually lead to
integrability being more ubiquitous than in higher dimensions),  numerical simulations have found
conflicting results concerning the existence of thermalization~\cite{manmana07_quench_spinless_fermions_nnn,kollath07_quench_BH,roux08_time_dependent_lanczos}. Indeed, in a recent work,  Rigol and
coworkers~\cite{rigol09_breakdown_thermalization_finite,santos10_quantum_chaos_1D} has also pointed out that the statistics of the constituent particles may also play an important role
in determining the thermalization dynamics.

Furthermore, in addition to studying the quench dynamics starting from a pure state, it can be also interesting to consider quantum quenches where
the system is initially prepared in a thermal state by being at $t < 0$ in contact with an energy reservoir at temperature $T$. After the quench, the system is  isolated from the reservoir and evolves unitarily according to the dynamics dictated by  different Hamiltonian $H_\text{f}$. This is the simplest kind of mixed initial state one can consider and it allows to analyze the effect on the quench dynamics of mixing (by means of the initial temperature) a fraction of the excited states with the ground state  of $H_\text{i}$ .

In this work, we shall analyze the quench dynamics of the sine-Gordon model (sGM). This is an integrable field theory,  but its exact solution is indeed quite difficult to deal with as the elementary excitations satisfy a rather non-standard algebra. Instead, we confine ourselves to two limits in which the model can  written as a quadratic Hamiltonian. In one of these two limits, the so called Luther-Emery (LE) limit, using a trick called refermionization,~\cite{giamarchi04_book_1d,gogolin98_1D_book} the model  describes a system of one-dimensional massive (\emph{i.e.} gapped) Dirac fermions. The fermions and their anti-particles (or holes, to use a solid-state Physics language)  describe the solitonic and anti-solitonic excitations of the sGM, which, in the LE limit, happen to be non-interacting. In another limit,  the model can be rather well approximated by a quadratic model of massive bosons. The latter describe a series of bound states of solitons and anti-solitons in the limit where interaction between them is strongly attractive. As explained elsewhere,~\cite{iucci09_quench_LL,cazalilla06_quench_LL} the quench dynamics of a quadratic Hamiltonian can be solved exactly via a time-dependent Bogoliubov transformation. We shall consider here two kinds of quenches, which correspond to the appearance and disappearance of the mass term (\emph{i.e.} the gap) in the sGM. At zero temperature, we found correlations when the system is quenched from the gapped to the gapless phase at zero or low temperature exhibit an exponential decay with a correlation length/time fixed by the gap. This result is in agreement with the  results of Calabrese and Cardy,~\cite{calabrese06_quench_CFT,calabrese07_quench_CFT_long} based on a mapping to a boundary conformal field theory. At high temperatures, however, the correlation length (time) is determined by the temperature. At intermediate temperatures, the system will exhibit a crossover between the zero-temperature (gap dominated) and the high-temperature (temperature dominated) regimes. On the other hand, correlations following a  quench from the gapped to the gapless phase at the Luther-Emery and the semi-classical limit exhibit a somewhat different behavior, which may indicate a break-down of the semiclassical approximation or a qualitative change in the behavior of correlations as one moves away from the Luther-Emery limit.

This paper is organized as follows. In section \ref{sec:sG}, we  introduce the (quantum) sine-Gordon mode, its phases,  and describe the problem of quantum quenches in this model. In section~\ref{sect:LE}, we consider quenches in the Luther-Emery limit at zero temperature, whereas in section~\ref{sect:SC} we take up the quenches in semiclassical limit. In section~\ref{sec:finiteT}, we consider the effect of thermal fluctuations in the initial state assuming that contact with the energy reservoir is removed at the time when the system is quenched and therefore, it subsequently evolves in isolation. In section~\ref{sec:generalized}, we take up the issue of the long-time asymptotic behavior of correlations and expectation values, and show that it is given by a generalization of the Gibbs ensemble, as recently pointed out by Rigol and coworkers.~\cite{rigol07_generalized_gibbs_hcbosons} Finally in section~\ref{sec:exp} we discuss some possible experimental consequences of this work. We provide a summary of the conclusions in section~\ref{sec:conclusions}. The results for the correlations at zero temperature that are reported in this manuscript have been available previously as a part of an unpublished preprint.~\cite{iucci09_quench_preprint} Since our preprint appeared other authors have also considered the thermalization dynamics in the sine-Gordon model.
In particular (see this special issue),  Sabio and Kehrein~\cite{sabio10_quench_sine_gordon} used a flow equation method. Fioretto and Mussardo used~\cite{fioretto10_quench_integrable_theories}  form factor methods to tackle quantum quenches in the sine-Gordon model, finding strong evidence that  the long time behavior of a local operator is described by a generalized Gibbs ensemble.   Furthermore, using the time-dependent renormalization group~\cite{cazalilla02_tDMRG,vidal98_simulations_1D_lattice},  Barmettler and coworkers~\cite{barmettler10_quench_heisenberg} have investigated the quench dynamics of the XXZ model, which in the continuum limit reduces to the sine-Gordon model

\section{The sine-Gordon model}\label{sec:sG}

The sine-Gordon model is described by the following Hamiltonian:
\begin{align}
H_\text{sG}(t) &= H_0 -  \frac{ \hbar v g(t)}{\pi a^2_0} \int dx \,\cos 2 \phi,\label{eq:sG}\\
H_0 &= \frac{\hbar v}{2\pi} \int  dx \,  : K^{-1}\left( \partial_x\phi\right)^2 + K \left(\partial_x \theta \right)^2 :\,,
\end{align}
where $:\ldots:$ stands for normal order of the operators,~\cite{haldane81_luttinger_liquid,giamarchi04_book_1d,gogolin98_1D_book} $a_0$ is
a short-distance cut-off,  and the phase and density fields, $\theta(x)$ and
$\phi(x)$, are canonically conjugated in the sense that the obey
$[\phi(x), \partial_{x'} \theta(x')] = i \pi \delta(x-x')$. This model can be regarded
as a perturbation the Luttinger model (see \emph{e.g.} Refs.~\onlinecite{haldane81_luttinger_liquid,giamarchi04_book_1d,gogolin98_1D_book} for a review),
which still yields an integrable model. In equilibrium the model
is known to exhibit two phases, which, according to the renormalization
group analysis,~\cite{giamarchi04_book_1d,gogolin98_1D_book} and for infinitesimal
and positive values of the coupling  in front of the cosine term, roughly correspond
to $K < 2$ (gapped phase) and $K \geq 2$  (gapless phase).

In order to study the non-equilibrium (quench)
dynamics, we will consider two different types of quenches:
the quench from  the gapless to the gapped phase and the reversed process,
from the gapless to the gapped. In the
first case, we assume that the \emph{dimensionless} coupling
$g(t)$ is suddenly turned on, \emph{i.e.} $g(t) = g\, \theta(-t)$.
With this choice, $H_\text{i} = H_\text{sG}(t \leq 0)$ is a
Hamiltonian whose ground state exhibits a frequency  gap, $m$, to all
excitations, whereas $H_\text{f} = H_\text{sG}(t > 0)$ has gapless
excitations. Conversely, in the second case, we consider
that $g(t)$ is suddenly turned off, \emph{i.e.}
$g(t)=g\,\theta(t)$. In this case, the ground state of
$H_\text{i}$ is gapless whereas the Hamiltonian performing the
time evolution, $H_\text{f}$, has gapped
excitations. However, although both $H_\text{i}$ and $H_\text{f}$
define integrable field theories, for a
general choice of the parameters $K$ and $g$, the quench dynamics
cannot be analyzed, in general,  by the elementary methods of Refs.~\cite{cazalilla06_quench_LL,iucci09_quench_LL}. Nevertheless, in two limits,
the Luther-Emery  (which corresponds to
$K = 1$, see Sect.~\ref{sect:LE})
and in the semiclassical (that is, for $K\ll 1$, Sect.~\ref{sect:SC}) limits, it is possible to study  the quench
dynamics by methods similar to those of  Ref.~\onlinecite{iucci09_quench_LL}. However, as explained above,
the statistics  of the elementary
excitations happens to be different in these two limits.

\section{The Luther-Emery limit}\label{sect:LE}

\subsection{Introductory remarks}

Let us start by considering the sine-Gordon model, Eq.~(\ref{eq:sG2}),
for $K = 1$, which is the so-called Luther-Emery limit. It is convenient
to introduce  rescaled density and phase  fields, which will be
denoted  as $\varphi(x) = K^{-1/2} \phi(x)$ and $\tilde{\varphi}(x)
=  K^{1/2} \theta(x)$. Thus, the Hamiltonian of Eq. (\ref{eq:sG}) becomes:
\begin{equation}
H_\text{sG}(t)= \frac{\hbar v}{2\pi} \int dx :\left(\partial_x \varphi\right)^2 +
\left(\partial_x \tilde{\varphi}\right)^2: - \frac{\hbar v g(t)}{\pi a_0^2} \int dx \, \cos  \kappa \varphi,\label{eq:sG2}
\end{equation}
where $\kappa = 2 \sqrt{K}$. At the Luther-Emery limit $\kappa = 2$ (\emph{i.e.} $K = 1$)
and the model can be rewritten as a one-dimensional model of  \emph{massive} Dirac  fermions with  mass
by using the following bosonization formula for the Fermi field operators,~\cite{giamarchi04_book_1d,gogolin98_1D_book}
\begin{equation}\label{eq:bosonization_formula}
\psi_\alpha(x)=\frac{\eta_\alpha e^{is_{\alpha} \pi/4}}{\sqrt{2\pi a}}\, e^{i
s_\alpha\phi_\alpha(x)},
\end{equation}
where $s_{r} = - s_{l}= +1$ and  the chiral fields  $\phi_r(x) = \varphi(x) + \tilde{\varphi}(x) $, and
$\phi_l(x) =    \varphi(x) - \tilde{\varphi}(x) $. For
computational convenience, we  choose the Majorana fermions in
Eq.~(\ref{eq:bosonization_formula})   to be $\eta_{r} = \sigma_x$ and $\eta_{l} = i \sigma_y$, where
$\sigma_x$ and $\sigma_y$ are the familiar Pauli matrices.
In addition, we note that the gradient terms in Eq.~(\ref{eq:sG2}) can be written as the kinetic
energy of free massless Dirac fermions in one dimension:~\cite{haldane81_luttinger_liquid,giamarchi04_book_1d,gogolin98_1D_book}
\begin{equation}
H_{0} =  - i \hbar v \int dx  :\psi^{\dag}_r(x) \partial_x
\psi_r(x) - \psi^{\dag}_l(x) \partial_x \psi_l(x): \, .
\end{equation}
As far as the cosine operator is concerned,  the bosonization formula,
Eq.~(\ref{eq:bosonization_formula}), implies that:
\begin{align}
\psi^{\dag}_r(x) \psi_l(x) + \psi^{\dag}_l(x) \psi_r(x) &=
\frac{\Gamma}{\pi a_0}  \cos 2 \varphi(x) \label{eq:fermionmassterm1}\\
& = \frac{\Gamma}{\pi} :\cos 2\varphi(x) : \label{eq:fermionmassterm2}
\end{align}
where $\Gamma = i \sigma_x \sigma_y$.   This is almost the
cosine term  of the sGM in the LE limit (cf. Eq.~\ref{eq:sG2}),
except for the presence of the operator $\Gamma$.
However, we note that   $\Gamma^2 = 1$ and that this operator also
commutes with  $H_0$ and  with operator in the left hand-side
of  Eq.~(\ref{eq:fermionmassterm1}). The  first property implies that the eigenvalues of $\Gamma$ are $\pm 1$ whereas  the second property implies that
$H_\text{LE}(t) = H_0 +  \hbar v g(t) \int dx \left[ \psi^{\dag}_r(x) \psi_l(x) + \psi^{\dag}_l(x) \psi_r(x) \right] $
and $\Gamma$ can be diagonalized simultaneously. Upon choosing the eigenspace where $\Gamma = -1$, we obtain that
\begin{align}
H_\text{LE}(t) &=  - i \hbar v \int dx :\psi^{\dag}_r(x)\partial_x\psi_r(x)-\psi^{\dag}_l(x)\partial_x\psi_l(x): \nonumber \\
& \quad + \hbar v g(t) \int dx\,\left[ \psi^{\dag}_r(x) \psi_l(x) + \psi^{\dag}_l(x) \psi_r(x) \right], \label{eq:LEham}
\end{align}
is equivalent to Eq.~(\ref{eq:sG2}) when $\kappa = 2$.

To gain some insight into the phases described by the sGM,
let us  first consider the Luther-Emery Hamiltonian, $H_\text{LE}$
in two (time-independent) situations:  \emph{i}) $g(t) = 0$ (the gapless free
fermion phase, which  coincides with the Luttinger model for $K = 1$~\cite{haldane81_luttinger_liquid,giamarchi04_book_1d,gogolin98_1D_book}),
and  \emph{ii}) $g(t) = g > 0 $, \emph{i.e.}
a time-independent constant (which corresponds to the gapped phase). In order
to diagonalize the Hamiltonian, it is convenient  to work in
Fourier space and  write the fermion field operator as:
\begin{equation}
\psi_{\alpha}(x) = \frac{1}{L} \sum_p  e^{-a_0 |p|} \, e^{i p x} \,
\psi_{\alpha}(p). \label{eq:fermifourier}
\end{equation}
where $\alpha=r,l$. The limit where the cut-off  $a_0 \to 0^{+}$ should be formally
taken at the end of the calculations, but in some cases we shall not do it
in order to regularize certain short-distance divergences of the quantum sGM.
It will also useful to introduce a spinor whose components are the right and
left moving fields, and which will make the notation more compact:
\begin{align}
\Psi(p) = \begin{bmatrix} \psi_r(p) \\ \psi_l(p)
\end{bmatrix}, \quad \mathcal{H}(p) =  \begin{bmatrix}\hbar \omega_0(p) & 0\\
0 & -\hbar \omega_0(p)
\end{bmatrix},
\end{align}
Thus the Hamiltonian for the gapless phase, $H_0$, reads:
\begin{equation}
H_0 = \sum_p:\Psi^\dagger(p)\cdot\mathcal{H}_0(p)\cdot\Psi(p):\,, \label{eq:h0le}
\end{equation}
where $\omega_0(p)=vp$ is the fermion dispersion. However, the Hamiltonian
of the gapped phase, corresponding to $g(t) = g > 0$,  $H_\text{LE}$,   is not diagonal
in terms of the right and left moving Fermi fields. In the compact spinor notation
it reads:
\begin{align}
H_\text{LE} =  \sum_p :\Psi^{\dag}(p) \cdot \mathcal{H}(p)
\cdot \Psi(p):,\label{eq:H_fermionic}
\end{align}
where
\begin{align}
\mathcal{H}(p) =  \begin{bmatrix}\hbar \omega_0(p) &
\hbar m\\
\hbar m & -\hbar \omega_0(p)
\end{bmatrix},\label{eq:H_fermionic_matrix}
\end{align}
being $m = v g$. Nevertheless, $H_\text{LE}$ can be rendered diagonal
by means of the  following  unitary transformation:
\begin{equation}
\tilde{\Psi}(p) = \begin{bmatrix} \psi_{c}(p) \\
\psi_v(p)
\end{bmatrix} =  \begin{bmatrix} \cos \theta(p) &
\sin \theta(p) \\ -\sin \theta(p) & \cos \theta(p)
\end{bmatrix} \begin{bmatrix}\psi_r(p) \\ \psi_l(p)
\end{bmatrix}, \label{eq:trans}
\end{equation}
being
\begin{equation}
\tan2\theta(p) = \frac{m}{\omega_0(p)}.
\end{equation}
Thus the Hamiltonian of the gapped phase, in diagonal form, reads (after dropping
an unimportant  constant that amounts to the ground state energy):
\begin{equation}
H_\text{LE} = \sum_p \hbar \omega(p) \left[ :\psi^{\dag}_c(p) \psi_c(p) -
\psi^{\dag}_v(p) \psi_v(p): \right]. \label{hgapped}
\end{equation}
where $\omega(p) = \sqrt{\omega_0(p)^2 + m^2}$. We associate
$\psi^{\dag}_v(p)$ ($\psi^{\dag}_c(p)$) with the creation operator
for particles in the valence (conduction) band.

Before considering quantum quenches,  let us briefly discuss some of the
the properties of the ground states of the Hamiltonians
$H_0$ and $H_\text{LE}$. In what follows, these states will be
denoted  as  $| \Phi_0\rangle$ and $| \Phi \rangle$, respectively.
As mentioned above, the spectrum of $H_0$  is gapless, and the fermion
occupancies in the ground state $|\Phi_0\rangle$ are:
\begin{align}
n_r(p) &= \langle \Phi_0| \psi^{\dag}_r(p) \psi_r(p) |
\Phi_0 \rangle = \theta(-p), \label{eq:gs11}  \\
n_l(p)  &= \langle \Phi_0| \psi^{\dag}_l(p) \psi_l(p) |
\Phi_0 \rangle = \theta(p). \label{eq:gs12}
\end{align}
That is, all single-particle levels with negative momentum are filled out. However, $H_\text{LE}$ has a gapped spectrum and, therefore, when constructing its ground state, $|\Phi\rangle$, only the levels in the valence band  (which have negative energy) are filled, whereas the levels in the conduction band remain empty:
\begin{align}
n_v(p) &= \langle \Phi| \psi^{\dag}_v(p) \psi_v(p) |
\Phi\rangle = 1 \label{eq:occv},\\ n_c(p) &= \langle \Phi|
\psi^{\dag}_c(p) \psi_c(p) | \Phi \rangle = 0. \label{eq:occc}
\end{align}

\subsection{Quench from the gapped to the gapless phase}\label{sec:quench1}

The first situation we shall consider is when $g(t) = g\: \theta(-t)$ in Eq.~(\ref{eq:LEham}), so that the spectrum of the Hamiltonian abruptly changes from gapped to gapless (\emph{i.e.} quantum critical). In the following we denote $H_\text{i}\equiv H_\text{LE}(t < 0) = H_\text{LE}$ and $H_\text{f}\equiv H_\text{LE}(t > 0) = H_0$. The time evolution of $\psi_{r,l}$ for $t > 0$ is thus
\begin{equation}\label{eq_time_evolution_rl_gapped_to_gappless}
\psi_{r,l}(p,t)=e^{\pm i\omega_0(p)t}\psi_{r,l}(p).
\end{equation}
We first consider the zero temperature quench in this section, and postpone to Section~\ref{sec:finiteT} the discussion of the more complicated finite-temperature case. In the zero temperature case the initial state $\rho_{\rm i} = |\Phi\rangle \langle \Phi|$. Notice that,  in this state, $\langle \Phi |  \cos 2 \phi(x) |\Phi \rangle = \langle \Phi |  \cos 2 \varphi(x) |\Phi \rangle = \Re\langle \Phi|  e^{-2 i\varphi(x)}| \Phi \rangle  =  - \langle \psi^{\dag}_{r}(x) \psi_{l}(x) \rangle \neq  0$ (the minus sign stems from the eigenvalue of the operator $\Gamma  = \eta_r \eta_l$), whereas in the ground state of $H_0$, $|\Phi_0\rangle$, the expectation value of the same operator vanishes. Therefore,  it behaves like an order parameter in equilibrium, and we can expect that it exhibits interesting dynamics out of equilibrium. Indeed,
\begin{equation}
\mathcal{C}(t) =\langle  e^{-2 i \varphi(x,t)} \rangle = - \frac{1}{L} \sum_p \langle \psi^{\dag}_r(p,t) \psi_l(p,t)\rangle =\frac{1}{2L}  \sum_p e^{-2i\omega_0(p)t} \, \sin 2\theta(p),
\end{equation}
where, to evaluate the expectation value  we have set $T  =  0$  and therefore $\langle\psi^{\dag}_r(p) \psi_l(p)\rangle = -\frac{1}{2} \sin 2\theta(p)$, as it follows from Eqs.~(\ref{eq:trans},\ref{eq:occv},\ref{eq:occc}).
The above expression can be readily evaluated by recalling that
$\sin 2\theta(p) = m/\omega(p)$, which
yields, in the $L \to \infty$ limit,
\begin{equation}
\mathcal{C}(t)  =  m \int^{+\infty}_{0}\frac{dp}{2\pi} \frac{\cos2 \omega_0(p) t}{\omega(p)} = \left(\frac{m}{2\pi v} \right) K_0\left(2m t\right) \simeq\frac{1}{4v}\sqrt{\frac{m}{\pi t}} e^{-2 m t},\label{eq:order_parameter_0T_gapped_to_gappless}
\end{equation}
where $K_0$ is the modified Bessel function.
Thus we see that the `order parameter' $\langle \cos 2 \varphi(0,t)\rangle$ decays
exponentially at long times at $T = 0$. The decay rate is proportional to the
gap between the ground state (the initial state) and the  first excited state of the
initial Hamiltonian $H_{\rm i} = H$. The existence of this gap means, in particular,
that correlations in the initial state between  degrees of freedom of the system are exponentially suppressed
beyond a distance of the order $\xi_c \approx v/m$. Upon quenching, the  evolution of the  system is
dictated by a critical  Hamiltonian, $H_0$, that is, a Hamiltonian  describing
excitations that propagate ballistically along `light cones'  corresponding to the `trajectories'
$x \pm v t$. Thus, as discussed by  Calabrese and Cardy~\cite{calabrese06_quench_CFT,calabrese07_quench_CFT_long}
the correlation length scale characterizing the initial state translates  into an exponential
decay in time of the order parameter at long times. This exponential decay  is also found  (for the same type of quench) in the semiclassical  limit of the sine-Gordon model (see Sect.~\ref{sect:SC} below).

Next we shall consider the (equal-time) two-point correlation function of the same operator, namely:
\begin{equation}
\mathcal{G}(x,t) = \langle e^{-2i \varphi(x,t)} e^{2i \varphi(0,t)}\rangle.
\end{equation}
When using the fermionic representation of $e^{2i\varphi(x,t)}$, and upon expanding  in Fourier modes
we arrive at:
\begin{equation}
\mathcal{G}(x,t) =
\frac{1}{L^2} \sum_{p_1,p_2,p_3,p_4} e^{i(p_1-p_2)x}e^{-i[\omega_0(p_1) + \omega_0(p_2) - \omega_0(p_3) -
\omega_0(p_4)]t}\langle \psi^{\dag}_r(p_1) \psi_l(p_2) \psi^{\dag}_l(p_3)\psi_r(p_4)\rangle. \label{eq:twopointle2}
\end{equation}
Applying Wick's theorem, there are three different contractions of the
above four fermion expectation value, which can be evaluated using
Eqs.~(\ref{eq:trans},\ref{eq:occv},\ref{eq:occc}). This yields the following
contractions:
\begin{align}
\langle \psi^{\dag}_r(p) \psi^{\dag}_l(p)\rangle &= \langle
\psi_r(p) \psi_l(p)\rangle = 0\label{eq:corr_dagger}\\
\langle \psi^{\dag}_l(p) \psi_r(p) \rangle &= \langle \psi^{\dag}_r(p)
\psi_l(p) \rangle = - \frac{1}{2} \sin 2\theta(p),\label{eq:corr_rl} \\
\langle \psi^{\dag}_r(p) \psi_r(p) \rangle &= \sin^2 \theta(p), \label{eq:corr_r}\\
\langle \psi_l(p) \psi^{\dag}_l(p) \rangle &= \cos^2 \theta(p).\label{eq:corr_l}
\end{align}
Hence, in the thermodynamic limit ($L\to \infty$), we obtain:
\begin{equation}\label{eq:two_points_LE}
\mathcal{G}(x,t) = \mathcal{C}(x,t)\mathcal{C}(0,t)+\vert\frac{i}{2}\delta(x)+\mathcal{H}(x,t)\vert^2
\end{equation}
where
\begin{equation}
\mathcal{H}(x,t)=\int_0^{+\infty}\frac{dp}{2\pi} \sin px\frac{\omega_0(p)}{\omega(p)}
e^{-a_0 p}= \frac{m}{2\pi v}K_1\left(\frac{m |x|}{v}\right).
\end{equation}
in the limit $a_0\to0$. Therefore, for $|x|\neq 0$,
\begin{equation}
\mathcal{G}(x,t) = \left(\frac{m}{2\pi v}\right)^2
\left(\left[K_0\left( 2 m t\right) \right]^2 +
\left[K_1\left(\frac{m |x|}{v}\right)\right]^2\right). \label{eq:twopointle3}
\end{equation}
Let us examine the behavior of this correlation function in the
asymptotic limit where $|x| \gg \xi_c= v/m$ and $2v t \gg \xi_c$.
Since the Bessel functions decay exponentially for large values of their arguments,
the leading term in   $\mathcal{G}(x,t)$ depends
on whether $t < |x|/2v$ or $t > |x|/2v $. Thus,
\begin{align}
\mathcal{G}(x,t)  \simeq
\begin{cases}
\frac{m}{16\pi v x} e^{-2m |x|/v} & t > |x|/2v,\\
\frac{m}{32\pi v^2 t} e^{-4mt}  & t < |x|/2v.
\end{cases}
\end{align}
These results are in also agreement with those obtained using a mapping to a  boundary
conformal field theory  by Calabrese and Cardy for general quantum quenches from a non-critical into a critical
state.\cite{calabrese06_quench_CFT,calabrese07_quench_CFT_long}

\subsection{Quench from the gapless to gapped phase}\label{sec:le_gaplesstogapped}

We next consider the reversed situation to the one discussed
in the previous subsection. In this case, we
set $g(t) = g \: \theta(t)$  in Eq.~(\ref{eq:LEham}), \emph{i.e.} the initial state is critical and
corresponds to the ground state of $H_\text{i}=H_0$, whereas
the time evolution is performed according to $H_\text{f}=H_\text{LE}$.  We shall
consider the same correlation functions as in the previous subsection and therefore it is
convenient in this case to obtain the time evolution of  the
operators $\psi_r(p)$ and $\psi_l(p)$, whose action on the initial
state is known [cf. e.g. Eqs.~(\ref{eq:gs11},\ref{eq:gs12})].
Once again, we first restrict ourselves to the $T = 0$
case and defer the discussion of finite temperature effects to
Sect.~\ref{sec:finiteT}. We first note that the time-evolved Fermi operators can be related to the
operators at $t = 0$ by means of the following (time-dependent) transformation:
\begin{align}
\psi_r(p,t) &= e^{i H_\text{f} t} \psi_r(p) e^{-iH_\text{f} t
/\hbar} =f(p,t) \psi_r(p) + g^{*}(p,t) \psi_l(p), \label{eq:psirt}\\
\psi_l(p,t)  &= e^{i H_\text{f} t} \psi_l(p) e^{-iH_\text{f} t
/\hbar} = g^*(p,t) \psi_r(p) + f^*(p,t) \psi_l(p),\label{eq:psilt}
\end{align}
where $f(p,t) = \cos \omega(p) t - i \cos 2\theta(p) \sin
\omega(p)t$ and $g(p,t) = i \sin 2\theta(p) \sin \omega(p) t$. This transformation can be
shown to respects the anti-commutation relations characteristic of Fermi statistics, and it
is therefore  a canonical transformation.
Using Eqs.~(\ref{eq:psirt}) to (\ref{eq:psilt}),
we can now compute the decay of the order parameter operator $e^{2i\varphi(x,t)}$. The calculation
yields:
\begin{equation}
\mathcal{C}(x,t) = -\frac{2}{L} \sum_{p >0}
\Re\left[  f^*(p,t) g(p,t)  \right].
\end{equation}
In deriving the above expression we have used that $f(-p,t) = f^*(p,t)$
and $g(-p,t) = g(p,t)$, which follows from
$\cos 2\theta(-p) = - \cos 2\theta(p)$ because $\cos 2\theta(p) =
\omega_0(p)/\omega(p)$. Thus, setting $ \Re \left[ f^*(p,t) g(p,t) \right] = - \cos
2 \theta(p) \sin 2\theta(p) \sin^2 \omega(p)t$ and taking
$L \to +\infty$, we find:
\begin{equation}
\mathcal{C}(x,t) = 2\int^{+\infty}_0
\frac{dp}{2\pi} \frac{m \omega_0(p)}{[\omega(p)]^2} e^{-a_0 p} \sin^2 \omega(p) t= A(m a_0)+\frac{m}{2\pi v}\ci(2mt),
\end{equation}
where $\ci$ is the cosine integral function. The first term is a non-universal constant that depends on the
short-distance cut-off $a_0$ introduced above (cf. Eq.~\ref{eq:fermifourier}). For long times this expression can be approximated by
\begin{equation}
\mathcal{C}(x,t)\simeq A(m a_0) + \frac{1}{4\pi v t} \sin 2 mt + O(t^{-2}).\label{eq:order_param_3}
\end{equation}
Hence we conclude that, when quenched from the critical (gapless)
phase into the gapped phase,  the order parameter exhibits
an oscillatory decay towards a (non-universal) constant value, $A( m a_0)$.

Using similar methods the  (equal-time) two-point correlation function,
$\mathcal{G}(x,t) = \langle e^{-2i \varphi(x,t)} e^{2i \varphi(0,t)}\rangle$,
can be also evaluated. The resulting expression  can be cast in a form identical
to Eq.(~\ref{eq:two_points_LE}) of subsect.~\ref{sec:quench1}. In the thermodynamic limit,
we find that, in the present case,  the function $\mathcal{H}(x,t)$ takes the form:
\begin{equation}\label{eq:calh}
\mathcal{H}(x,t)=-\int_{0}^{\infty}\frac{dp}{2\pi}\left\{-1+
\left[1-\cos2\omega(p)t\right]\frac{m^2}{[\omega(p)]^2}\right\}e^{-a_0 p}\: \sin px
\end{equation}
We have been unable to obtain an closed analytical expression for this function at all times. However,
in  $t\to +\infty$ limit, in which case the term in the integrand
proportional to $\cos2\omega(p) t$  oscillates very rapidly and therefore
yields a vanishing contribution, an analytical expression can be obtained.
Upon performing the momentum integral,
we obtain the following result for large $|x|$ (after taking the limit $a_0\to0$):
\begin{equation}
\mathcal{H}(x,t\to\infty)\approx \frac{-4v^2}{2\pi m^2|x|^3}.
\end{equation}
Hence, we obtain the following  asymptotic behavior
of the  two-point correlation  for $t\to \infty$:
\begin{equation}\label{eq:two_points_LE2}
\lim_{t \to +\infty}\mathcal{G}(x,t) = \left[ A(m a_0) \right]^2
+\frac{4v^4}{(2\pi)^2m^4x^6}.
\end{equation}
This result is clearly different from the equilibrium behavior of the same correlation function in the gapped phase,
where it decays exponentially to a constant.~\cite{giamarchi04_book_1d,gogolin98_1D_book} Instead, when the system
is quenched from the gapless into the gapped phase, we find that both the order parameter and the two-point correlation
function  (Eqs.~\ref{eq:order_param_3} and \ref{eq:two_points_LE2}) decay algebraically to  constant (non-universal)  values.

\section{The semiclassical limit}\label{sect:SC}

\subsection{Introductory remarks}

A good approximation to the sine-Gordon model (cf. Eq.~\ref{eq:sG2})
can be obtained  in the limit where $\kappa \ll 1$, which corresponds to
the $K\ll 1$ limit in the original notation of Eq.~(\ref{eq:sG}).
In this limit, we can expand the cosine term in~(\ref{eq:sG2})
about one of its minima,  \emph{e.g.}  $\varphi = 0$. Retaining
only the leading quadratic term  yields the following
\emph{quadratic} Hamiltonian for the boson field $\varphi(x)$:
\begin{equation}
H_\text{sG} \simeq H_\text{sc} = \frac{\hbar v}{2\pi} \, \int dx \left[
:\left(\partial_x \varphi(x)  \right)^2 + K \left(\partial_x
\tilde{\varphi}(x) \right)^2:\right] + \frac{\hbar v g(t) \kappa^2}{2\pi a^{2-\kappa^2/2}_0} \int dx \,
:\varphi^2(x):\,. \label{eq:hqcf}
\end{equation}
Within this approximation, the problem of studying a quantum quench in the
sine-Gordon model becomes  akin to the general problem studied in
Ref.~\onlinecite{iucci09_quench_LL}.  To see this, let us first expand $\varphi(x)$
in Fourier modes:
\begin{equation}
\varphi(x) = \frac{\phi_0}{\sqrt{K}} + i \frac{\pi x}{\sqrt{K}L}
\delta N  +  \frac{1}{2}\sum_{q\neq 0} \left( \frac{2\pi v}{\omega_0(q) L }\right)^{1/2}
\left[ e^{i q x} b(q) + e^{-iq x} b^{\dag}(q)
\right],\label{eq:mode_decomposition_varphi}
\end{equation}
where $\omega_0(q) = v |q|$; the $b$-operators introduced
above obey the standard Heisenberg algebra:
\begin{equation}
\left[ b(q), b^{\dag}(q') \right] = \delta_{q,q'},
\end{equation}
commuting otherwise. The first two terms in Eq.
(\ref{eq:mode_decomposition_varphi}) are the so-called zero-modes,
whose dynamics is only important at finite $L$. In what follows we
restrict our attention to the thermodynamic
limit ($L\to\infty$) and therefore neglect the dynamics of those
zero modes. Introducing~(\ref{eq:mode_decomposition_varphi})
into (\ref{eq:hqcf}), the Hamiltonian takes the general  form:
\begin{equation}
H(t) =\sum_{q} \hbar \left[  \omega_0(q) + m(q,t)
\right] b^\dag(q) b(q) + \frac{1}{2} \sum_{q} \hbar g(q,t) \left[ b(q) b( -q) +
b^\dag(q) b^\dag(-q)\right], \label{eq:genham}
\end{equation}
with the following identifications $\omega_0(q) = v |q|$ and $g(q,t)=m(q,t)=2 v
g(t)\kappa ^2/|q|a_0^{2-\kappa^2/2}$. As in the study of the Luther-Emery limit, we
shall assume that $g(t) = g \: \theta(-t)$, which corresponds to a quench from the gapped to a
gapless phase,\footnote{This case was studied  earlier by
Calabrese and Cardy in Ref.~\onlinecite{calabrese07_quench_CFT_long}, although not
as limit of the sine-Gordon model and therefore they considered the quench dynamics of different observables.}, or $g(t) = g\: \theta(t)$, which corresponds to a quench for the gapless to the gapped phase.
Following the procedure
described in the Appendix of Ref.~\onlinecite{iucci09_quench_LL},
the quench dynamics of this Hamiltonian can be solved by the following canonical transformation (indeed, the bosonic
version of Eqs.~\ref{eq:psirt}, \ref{eq:psilt}):
\begin{equation}
b(q,t) = f(q,t)\, b(q) + g^*(q,t) \,
b^\dag(-q),\label{eq:solution_time}
\end{equation}
where
\begin{align}
f(q,t) &= \cos \omega(q) t  - i \sin \omega(q) t \, \cosh
2\beta(q),
\label{eq:fqt}\\
g(q,t) &= i \sin \omega(q) t \, \sinh 2\beta(q).\label{eq:gqt}
\end{align}
Introducing $m^2=4 g v^2\kappa^2/a_0^{2-\kappa^2/2}$, which is the
the gap in the frequency spectrum of the gapped phase and
setting $m(q) = g(q)  =m^2/2\omega_0(q)$, the parameter $\beta(q)$  satisfies:
\begin{equation}
\tanh 2\beta(q)=\frac{m^2/2}{\omega^2_0(q)+m^2/2},
\label{eq:betaqq}
\end{equation}
and  the frequency:
\begin{equation}
\omega(q)=\sqrt{\omega_0(q)^2+m^2}. \label{eq:disperq}
\end{equation}
is the dispersion of the excitations in the gapped phase.

\subsection{Quench from the  gapped to the gapless phase}\label{sec:sc_gappedtogapless}

Let us begin by discussing  the situation where $g(t) = g \: \theta(-t)$.
In this case, the initial state is the ground state of the following Hamiltonian
(we omit the zero-mode part henceforth):
\begin{equation}
H_{\rm i} = H_{\rm sc} = \sum_{q\neq} \hbar \omega(q) \: a^{\dag}(q) a(q),
\end{equation}
where the operators $a(q)$ and $a^{\dag}(q)$ are bosonic operators related to
$b(q)$ and $b^{\dag}(q)$ by means of the following canonical transformation:
\begin{align}
a(q) & = \cosh \beta(q) \, b(q) + \sinh \beta(q) \, b^\dag(-q),
\label{eq:bogol}
\end{align}
with $\beta(q)$ satisfying Eq.~(\ref{eq:betaqq}).
At $t= 0$ the Hamiltonian abruptly  changes
to $H_{\rm f} = H_{0}$, which is diagonal in the $b(q)$ and $b^{\dag}(q)$ basis,
namely,
\begin{equation}
H_0 = \sum_{q\neq 0} \hbar \omega_0(q) \: b^{\dag}(q) b(q).
\end{equation}

 In this case
the evolution of the expectation value of the
order parameter operator $e^{-2i\phi(x)} = e^{-i\kappa \varphi(x) }$
or its correlation functions can be obtained from the knowledge of
the two-point (equal time) correlation function out of equilibrium
for the boson field $\varphi(x)$, \emph{i.e.}
$\mathcal{F}(x,t) = \langle\varphi(x,t) \varphi(0,t)\rangle - \langle \varphi^2(0,t)\rangle$, where
the expectation value is taken over the ground state of $H_{\rm sc}$ but the time
evolution is dictated by $H_0$. To compute this
object,  we first insert into the expectation value the Fourier expansion of  $\varphi(x)$,
Eq.~(\ref{eq:mode_decomposition_varphi})  and use Eq.~(\ref{eq:solution_time}).
Thus, we arrive at:
\begin{equation}
\langle \varphi(x,t) \varphi(0,t) \rangle = -\frac{1}{4}
\sum_{q\neq 0} \left( \frac{2\pi v}{\omega_0(q) L} \right) \Big[   \sinh 2\beta(q)  \cos \left( qx  - 2\omega_0(q) t \right)
 - e^{i q x} \sinh^2 \beta(q) - e^{-i qx} \cosh^2\beta(q)
\Big].\label{eq:propphi}
\end{equation}
Using this result, let us consider the behavior
of the order parameter following the quench. Taking into account that $\langle
e^{-2i \phi(0,t)} \rangle = \langle e^{-i \kappa \varphi(0,t)}
\rangle = e^{-\frac{\kappa^2}{2} \langle \varphi^2(0,t)\rangle}$,
we see that $\langle \varphi^2(0,t)\rangle$ must be evaluated in closed
form using Eq.~(\ref{eq:propphi}) . Before performing any
manipulation of this expression, it is convenient to subtract the constant $\langle
\varphi^2(0,0)\rangle$, which is formally infinite (\emph{i.e.} it depends
on the short distance cut-off, $a_0$). Thus, taking the thermodynamic limit where $L\to\infty$,
we obtain:
\begin{equation}
\langle \varphi^2(0,t) \rangle  - \langle \varphi^2(0,0)   \rangle = \frac{1}{2}\int^{+\infty}_0 \frac{d (q v)}{\omega(q)}\, \left[  \left(\frac{\omega(q)}{\omega_0(q)} \right)^2 - 1 \right]\sin^2 \omega_0(q) t. \label{eq:phi2_1}
\end{equation}
Inserting the expressions for $\omega(p)$ and $\omega_0(p)$ in the
above equation, we obtain:
\begin{equation}
\langle \varphi^2(0,t) \rangle - \langle \varphi^2(0,0)   \rangle  = -f(2mt/\hbar),
\end{equation}
where $f(z)$ is defined as:
\begin{equation}
f(z)= 1+\frac{1}{2}\,G^{21}_{13}\left(\frac{z^2}{4}\left|
\begin{matrix}
&  3/2 & \\
0 &1 &1/2
\end{matrix}\right.
\right),\label{eq:fz}
\end{equation}
being $G_{13}^{21}$ the Meijer $G$
function.\cite{gradshteyn_tables}
Using the asymptotic expansion for this function,
$f(z)\approx 1-\frac{\pi |z|}{2}$,  and hence the long-time
behavior of the order parameter is:
\begin{equation}
\langle e^{-2i\phi(0,t)}\rangle=  \langle e^{-2i\kappa \varphi(0,t)}\rangle = \langle e^{2i\phi(0,0)}\rangle
e^{\frac{\kappa^2}{8}(1-\pi m t)}. \label{eq:orderparamsc}
\end{equation}

We next examine the behavior of the two-point correlation function
of the same (order-parameter) operator,
\begin{equation}
\mathcal{G}(x,t) = \langle e^{2i\phi(x,t)}e^{-2i\phi(0,t)}\rangle =e^{\frac{\kappa^2}{2}  \mathcal{F}(x,t)},
\end{equation}
where we have defined $\mathcal{F}(x,t) = \langle \varphi(x,t) \varphi(0,t)\rangle
- \langle \varphi^2(0,t) \rangle$. At zero temperature, with the help of  Eq. (\ref{eq:propphi}),
we find that
\begin{equation}
\mathcal{F}(x,t)-\mathcal{F}(x,0)= -\frac{m^2}{4}\int_0^\infty\frac{d (v q)}{[\omega_0(q)]^2\omega(q)}(1-\cos qx)(1-\cos 2\omega_0(q)t),\label{eq:cc_mslmsv}
\end{equation}
where
\begin{equation}
\mathcal{F}(x,0)= -\frac{1}{2}\int_0^\infty\frac{d(v q)}{\omega(q)}\left(1-\cos
qx\right)e^{-q a_0}, \label{eq:cx0}
\end{equation}
being $a_0$ is the short-distance cut-off. Evaluating
the integrals:
\begin{equation}
\mathcal{F}(x,t)-\mathcal{F}(x,0)=f(mx/v)+f(2mt)-\frac{f[m(x/v+2t)]+f[m(x/v-2t)]}{2},
\end{equation}
where $f(z)$ has been defined in Eq.~(\ref{eq:fz}).
Thus, asymptotically (for $\max\{|x|/2v,t\} \gg m^{-1}$):
\begin{equation}
\mathcal{G}(x,t)={  e^{\kappa^2}  \cal G}(x,0) \times\begin{cases}
e^{- \kappa^2 \pi m     |x|/2v},  \text{for }t> |x|/2v \\
e^{- \kappa^2 \pi m t},  \text{for }t<|x|/2v, \label{eq:gxtsc}
\end{cases}
\end{equation}
where $\mathcal{G}(x,0)$ describes the correlations in the initial (gapped ground)
state, and exhibits the following asymptotic behavior:
\begin{equation}
\mathcal{G}(x,0)\simeq B(a_0) \left(1-\kappa^2\frac{\pi v}{m x}e^{-m|x|/v}\right)
\end{equation}
where $B(a_0)$ is a non-universal prefactor. Thus we see that
the asymptotic form of $\mathcal{G}(x,t)$ (Eq.~\ref{eq:gxtsc}),
as well that of the order parameter, Eq.~(\ref{eq:orderparamsc}), have the same form as the results
found limit the Luther-Emery limit, and also agree
with the  results of Calabrese and
Cardy based on a mapping to  boundary conformal field
theory.~\cite{calabrese06_quench_CFT,calabrese07_quench_CFT_long}

\subsection{Quench from the gapless to the gapped phase}\label{sec:gaplesstogapped}

In this case, the system finds itself initially in the ground state
of $H_\text{i}=H_0$, and suddenly (at $t = 0$)  the Hamiltonian is changed to $H_\text{f}=H_{\rm sc}$.
For this situation convenient to obtain the evolution of the
observables from the time-dependent canonical transformation
of Eq.~(\ref{eq:solution_time}), where  $\beta(q)$
and $\omega(q)$ are given by Eq.~(\ref{eq:betaqq}) and Eq.~(\ref{eq:disperq}),
respectively. In this case,
\begin{equation}
\langle \varphi(x,t) \varphi(0,t) \rangle = \frac{1}{4}
\sum_{q\neq 0} \left( \frac{2\pi v}{\omega(q) L} \right) \Big[   \sinh 2\beta(q) \cos \left( qx  - 2\omega(q) t \right)  +e^{i q x} \sinh^2 \beta(q) + e^{-i qx} \cosh^2\beta(q)
\Big]\,, \label{eq:propphi2}
\end{equation}
As in the previous subsection, $\langle e^{-2i \phi(x)} \rangle =
e^{-\frac{\kappa^2}{2} \langle \varphi^2(0,t) \rangle}$, and using
Eq.~(\ref{eq:propphi2}), we find that:
\begin{equation}
\langle \varphi^2(0,t)\rangle - \langle \varphi^2(0,0)\rangle =
\frac{1}{2}\int_0^\infty\frac{d (vq)}{\omega_0(q)}
\left[ \left( \frac{\omega_0(q)}{\omega(q)} \right)^2  -1 \right]
\sin^2\omega(q)t
\end{equation}
Note, interestingly, that this result can be obtained from Eq.~(\ref{eq:phi2_1}) by exchanging  $\omega_0(q)$ and $\omega(q)$.
However, when evaluating the integral we find that $\langle \varphi^2(0,t) \rangle  = +\infty$, for all $t > 0$,
due to the presence of infrared divergences that  are not cured by the existence  of a gap in
the spectrum of $H_{\rm f} = H_{\rm sc}$. Thus, we   conclude that  $\langle e^{-2i \phi(x)} \rangle =
e^{-\frac{\kappa^2}{2} \langle \varphi^2(0,t) \rangle}$ vanishes at all  $t > 0$.

The above result for the evolution of the order parameter seems to indicate
that the system apparently remains critical after the quench.
This is conclusion is also supported by the behavior of
the two-point correlation function of  the operator $e^{2i\phi(x,t)}$: Let $\mathcal{G}(x,t) =
\langle e^{2 i \phi(x,t) } e^{-2i \phi(0,t)} \rangle = e^{\kappa^2 \mathcal{F}(x,t)}$,
where $\mathcal{F}(x,t) = \langle \varphi(x,t)\varphi(0,t)\rangle - \langle \varphi^2(0,t)  \rangle$.
Using Eq.~(\ref{eq:solution_time}) and Eq.~(\ref{eq:mode_decomposition_varphi}),
we arrive at the following result (at zero temperature,  and for $L \to +\infty$):
\begin{equation}
\mathcal{F}(x,t)-\mathcal{F}(x,0)= \frac{m^2}{4}\int_0^\infty\frac{d(v q)}{\omega_0(q)\left[\omega(q)\right]^2}(1-\cos
qx)\left(1-\cos2\omega(q)t \right)
\label{eq:ctc0}
\end{equation}
where
\begin{equation}
\mathcal{F}(x,0)=-\int_0^\infty\frac{d(v q)}{2\omega_0(q)}(1-\cos
qx)e^{-q a_0}\label{eq:cc_msl_msv_inic}
\end{equation}
To illustrate the above point about the apparent ``criticality'' of the (asymptotic)
non-equilibrium state,we can analyze the
behavior of the two-point correlation function, $\mathcal{G}(x,t)$,
in two limiting cases, for $t=0$ and $t \to +\infty$.
At $t = 0$,  the correlation function, as obtained from
Eq.~(\ref{eq:cc_msl_msv_inic}), reads:
\begin{equation}
\mathcal{G}(x,0)= A(a_0) \: \left(\frac{a_0}{x}\right)^{2\kappa^2},
\end{equation}
where $A(a_0)$ depends on the short-distance cut-off $a_0$.
Thus, the correlations are power-law because the initial state is critical.
In the limit where $t\to+\infty$, the part of the integral
in Eq.~(\ref{eq:ctc0}) containing the  term $\cos 2\omega(q)t$  oscillates
very rapidly and upon integration averages to zero.
The remaining integral can be done with the help
of tables,\cite{gradshteyn_tables} yielding:
\begin{equation}
\mathcal{F}(x,t\to\infty)-\mathcal{F}(x,0)=-\frac{\sqrt{\pi}}{2}G^{22}_{04}\left(\frac{
m^2x^2}{4v^2}\left|
\begin{matrix}
& 1 & 1 & \\
1 & 1 & 0 & 1/2
\end{matrix}\right.
\right).
\end{equation}
where $G^{22}_{04}$ is a Meijer function Using
the asymptotic behavior of the Meijer function,\cite{gradshteyn_tables}
we obtain:
\begin{equation}
\lim_{t \to +\infty} \mathcal{G}(x,t)=B(a_0) \left(\frac{2v}{mx}\right)^{\kappa^2},
\end{equation}
with $B(a_0)$ being a non-universal constant.
Thus,  although initially the system is critical and therefore
correlations at equilibrium decay as a power law with exponent $2\kappa^2$,
when the system is quenched into a gapped
phase (where equilibrium correlations exhibit an exponential
decay characterized by a correlation length $\xi_c \approx v/m$),
the correlations remain power-law, within the semiclassical
approximation.  The exponent turns out to be smaller, equal to $\kappa^2$, which
is half the exponent in the initial (gapless) state.
In other words, within
this approximation, it seems that the system keeps
memory of its initial state, and behaves as if it was critical \emph{also}
after the quench.  This behavior seems somewhat different from the results
obtained for the same type of quench in the Luther-Emery limit, where both the order
parameter and the correlations for $t +\infty$ approach a constant value,  $A(ma_0)$ (unless the
non-universal amplitude $A(m a_0) =0$, which seems to require some fine-tuning).
Whether the differences found here between the Luther-Emery  and the semi-classical
limits are  due to a break-down of the quasi-classical
approximation, which neglects the existence of solitons and anti-solitons  in the spectrum of
the sine-Gordon model, or to a qualitative change in the dynamics as
one moves away from the Luther-Emery limit, it is not clear
at the moment. To clarify this issue will require further investigation with more sophisticated methods.

\section{Dynamics in the Luther-Emery limit at finite temperatures}\label{sec:finiteT}

In this section we shall consider that the initial state of the system corresponds to thermal mixed state, which describes a sGM system in contact with an energy reservoir (i.e. the canonical ensemble) at a temperature $T=\beta^{-1}$. The state is thus mathematically described by a density operator $\rho_\text{i} = e^{-H_\text{i}/T}/Z_\text{i}$.  We shall assume that the coupling to the thermal bath  is turned off at $t = 0$, and the system subsequently evolves unitarily in isolation according to $H_\text{f}$.
We shall  consider only the Luther-Emery limit of the sGM, where exact results can be obtained  at all temperatures within
the sGM model. The latter is not true in the semiclassical limit discussed above because this approximation
only captures the breather part of the spectrum and not the solitonic part. We shall therefore not consider finite temperature
quenches in this limit here.

\subsection{Quench from massive to massless}

Consider first the quench from the gapped to the gapless phase. The initial (gapped) Hamiltonian $H_\text{i}=H_\text{LE}$ is thus diagonal in the valence and conduction fermion basis and therefore immediately follows
\begin{align}
\langle\psi_{v}^{\dagger}(p)\psi_{v}(p')\rangle & =f_{\text{F}}[-\omega(p)]\delta_{pp'},\label{eq:averages_vv}\\
\langle\psi_{c}^{\dagger}(p)\psi_{c}(p')\rangle & =f_{\text{F}}[\omega(p)]\delta_{pp'},\label{eq:averages_cc}\\
\langle\psi_{v}^{\dagger}(p)\psi_{c}(p')\rangle & =0\label{eq:averages_cv}
\end{align}
Here $f_{\text{F}}$ is the Fermi factor and
\begin{equation}
\left\langle \cdots \right\rangle=\Tr\{ e^{-\beta H_{\text{i}}}\cdots\}.
\end{equation}
The final Hamiltonian $H_\text{f}=H_0$; thus, using Eqs. (\ref{eq:averages_vv})-(\ref{eq:averages_cv}) we can compute the finite temperature versions of Eqs.~(\ref{eq:corr_rl}) to (\ref{eq:corr_l}):
\begin{align}
\langle\psi_{R}^{\dagger}(p)\psi_{R}(p)\rangle &=\cos^{2}\theta_{p}f_{\text{F}}[\omega(p)]+\sin^{2}\theta_{p}f_{\text{F}}[-\omega(p)],\\
\langle\psi_{L}^{\dagger}(p)\psi_{L}(p)\rangle &=\sin^{2}\theta_{p}f_{\text{F}}[\omega(p)]+\cos^{2}\theta_{p}f_{\text{F}}[-\omega(p)],\\
\langle\psi_{R}^{\dagger}(p)\psi_{L}(p)\rangle &=\langle\psi_{L}^{\dagger}(p)\psi_{R}(p)\rangle =-\frac{1}{2}\sin2\theta_{p}\tanh\frac{\beta\omega(p)}{2}.
\end{align}
Note that these averages automatically vanish if the fermion operators are evaluated at different values of $p$ because of momentum conservation. The time evolution is again dictated by $H_{\text{f}} = H_0$ as in the zero-temperature case.

Let us next consider some interesting observables. We begin with the order parameter, which reads:
\begin{equation}
\mathcal{C}(t;\beta)=m\int_0^{\infty}\frac{dp}{2\pi} \frac{\cos2\omega_0(p)t}{\omega(p)}\tanh\frac{\hbar\beta\omega(p)}{2}.\label{eq:order_parameter_T_1}
\end{equation}
This integral can be transformed into an infinite sum by expanding $\tanh\beta\varepsilon/2$ in powers of $e^{-\beta\varepsilon}$ and integrating term by term. We thus obtain the following low temperature expansion:
\begin{equation}
\mathcal{C}(t;\beta)=\frac{m}{2\pi v}\sum_{n\in\mathbb{Z}}(-1)^{n}K_{0}\left[\sqrt{(2 m  t)^{2}+(n\hbar\beta m)^{2}}\right].
\end{equation}
Since the function $K_{0}$ decays exponentially for large value of its argument, from this expression we see that when the temperature is decreased, less terms are needed to  approximate the sum. In particular,  at zero temperature ($\beta\to \infty$) only the $n=0$ term contributes and we recover the zero temperature result of Eq. (\ref{eq:order_parameter_0T_gapped_to_gappless}). At finite but low temperatures, the asymptotic behavior at long times is an exponential decay where the characteristic time decay is fixed by the gap. However, at higher temperatures, more terms contribute to the sum whereas the alternating sign leads to some partial cancelations. As a result, we expect a faster decay in time of the order parameter. To further analyze the long time behavior in this regime, we shall use an identity of Bessel functions [see  Eq.~(\ref{eq:bessel_identity_1})]  which allows us to obtain the following high temperature expansion:
\begin{equation}
\mathcal{C}(t;\beta)=\frac{ m }{2v}\sum_{l\in\mathbb{Z}} \frac{\exp\left[-\frac{2 t}{\hbar\beta} \sqrt{\left(\hbar\beta m \right)^2+\pi^2(2l+1)^2}\right]}{\sqrt{\left(\hbar\beta m \right)^2+\pi^2(2l+1)^2}}.
\end{equation}
Note that, for this expansion, the higher the temperature the smaller the number terms that needs to be retained to accurately approximate the sum. In particular, the infinite temperature limit, $\beta m \hbar\ll1$, only the $l=0$ terms contributes and thus the decay in time is exponential,
\begin{equation}
\mathcal{C}(t;\beta)\approx\frac{ m }{2\pi v}\exp\left[-t/\tau_c\right],\quad  \hbar\beta m\ll1
\end{equation}
but now, the characteristic decay time $\tau_c$ is now fixed by the inverse temperature:
\begin{equation}
\tau_c=\frac{\hbar\beta}{2\pi}.
\end{equation}
Thus, to summarize, the asymptotic behavior of the order parameter following a quench from the gap to the gapless phase is described by  an exponential decay both at very low and very high temperatures.  At very low temperatures, the characteristic decay time is given by the frequency gap, $m$, but as  temperature of the initial state is increased,  the characteristic time is determined by the temperature. The behavior for intermediate temperatures is a crossover between these two exponentially decaying behaviors. Moreover, it is also worth mentioning that the exponential decay at large temperatures is characteristic of a critical theory at finite temperatures. Indeed, in the sGM we expect that, as the temperature is raised well above the gap energy scale, $\hbar m$, the
the properties of the system will become indistinguishable from those of a critical system.

 As for the the finite-temperature correlation function, it can be again cast in the same form as the zero temperature case, Eq.~(\ref{eq:two_points_LE}), with the function $\mathcal{H}(x,t;\beta)$ having the following $t \to +\infty$ limit:
\begin{equation}
\lim_{t\to+\infty}\mathcal{H}(x,t;\beta)=\int_0^{\infty}\frac{dp}{2\pi}
\frac{\omega_0(p)}{\omega(p)} \left[\tanh\frac{\hbar\beta\omega(p)}{2}\right]  \sin px
\end{equation}
Using the same technique as for the order parameter, we evaluate this function as
\begin{equation}
\lim_{t\to+\infty}\mathcal{H}(x,t;\beta) =\frac{mx}{2\pi v^2} \sum_{n=1}^{\infty}(-1)^n\frac{K_1\left(m\sqrt{\frac{x^2}{v^2}+n^2\hbar^2\beta^2}\right)} {\sqrt{\frac{x^2}{v^2}+n^2\hbar^2 \beta^2}}\label{eq:F_beta_x}
\end{equation}
The reasoning is the same as before: at very low temperatures, only the term with $n=0$ contributes, and we recover the zero temperature expression. However as the temperature increases more terms become important. We can use the dual expression Eq. (\ref{eq:bessel_identity_2}) which applied to Eq. (\ref{eq:F_beta_x}) yields
\begin{equation}
\lim_{t\to+\infty}\mathcal{H}(x,t;\beta) =\frac{1}{2v\hbar\beta}\sum_{l\in\mathbb{Z}} \exp\left[-\frac{x}{v\hbar\beta}\sqrt{(m\hbar\beta)^{2}+\pi^{2}(2l+1)^{2}}\right]
\end{equation}
This function decays exponentially for long distances (only the $l=0$ term contributes at high temperatures), with a characteristic correlation length that is fixed by the temperature of the initial state:
\begin{equation}
\lim_{t\to+\infty}\mathcal{H}(x,t;\beta) \simeq \frac{1}{2v\hbar\beta}e^{-|x|/\xi},\quad\xi=\frac{v\hbar\beta}{\pi}
\end{equation}
Upon introducing this results into (the finite temperature equivalent of) Eq.~(\ref{eq:two_points_LE}), we arrive at
an expression whose asymptotic behavior again depends on whether $|x| > 2vt$  or $|x| <  2 vt$ (with correlation
time/lengths given by $m$ or $T$ depending on the temperature range). Thus, the correlations at finite temperature
also will exhibit the so-called `light-cone' effect.~\cite{calabrese06_quench_CFT,calabrese07_quench_CFT_long}

\subsection{Quench from the gapless to the gapped phase}

We next consider that the system is quenched from the gapless phase intro the gapped phase.
Thus we need to assume that the system was initially described by $H_{\rm i} = H_\text{LE}$
and in contact with an energy reservoir at temperature  $T=\beta^{-1}$. The following expectation values (understood
over the initial thermal ensemble)  will be required in the calculations to follow:
\begin{align}
\langle\psi_{R}^{\dagger}(p)\psi_{R}(p)\rangle & =f_{\text{F}}[\omega_0(p)],\\
\langle\psi_{L}^{\dagger}(p)\psi_{L}(p)\rangle & =f_{\text{F}}[-\omega_0(p)],\\
\langle\psi_{R}^{\dagger}(p)\psi_{L}(p)\rangle & =\langle\psi_{L}^{\dagger}(p)\psi_{R}(p)\rangle=0\end{align}
and hence
\begin{align}
\langle\psi_{v}^{\dagger}(p)\psi_{v}(p)\rangle &=\frac{1}{2}\left(1+\cos2\theta_{p}\tanh\frac{\beta\omega_0(p)}{2}\right)\\
\langle\psi_{c}^{\dagger}(p)\psi_{c}(p)\rangle
&=\frac{1}{2}\left(1-\cos2\theta_{p}\tanh\frac{\beta\omega_0(p)}{2}\right)
\end{align}
Thus, the time evolution of the order parameter can be obtained and reads:
\begin{align}
\mathcal{C}(t;\beta)&=2\int_0^{\infty}\frac{dp}{2\pi}
 \frac{m\omega_0(p)}{\omega(p)^2}\tanh\frac{\hbar\beta\omega_0(p)}{2}e^{-a_0 p} \: \sin^2  \omega(p)t
\end{align}
As $t\to +\infty$ this function approaches a non-universal constant that depends on the energy cut-off $a_0$ and the temperature. At high temperatures,
\begin{equation}
\mathcal{C}(t,\beta)=A(ma_0,\beta)+\hbar\beta\frac{\sqrt{\pi m}}{16\pi v t^{3/2}}\sin\left(2 m t+\textstyle{\frac{\pi}{4}}\right)
\end{equation}
Thus, after the sudden quench at $t=0$ from a high temperature state in the critical regime into the gapped phase, the order parameter shows an oscillatory decay towards a constant value. However, the exponent of the decaying law is different from the decaying exponent in the case of a quench from a zero (or low) temperature state. The whole picture is the following: the order parameter exhibits an oscillatory decaying low as $t^{-3/2}$ for times smaller than a time scale fixed by the temperature, where there is a crossover to a $t^{-1/2}$ behavior characteristic of low temperatures.

 Concerning the two-point  correlation function, it can be again recast as in Eq.~(\ref{eq:two_points_LE}) with
 the function $\mathcal{H}(x,t;\beta)$ being given by:
\begin{equation}
\mathcal{H}(x,t;\beta)=-\int_0^{\infty}\frac{dp}{2\pi}\sin px \left\{-1+\left[1-\cos2\omega(p)t\right]\frac{m^2}{\omega(p)^2}\right\}\tanh\frac{\hbar\beta\omega_0(p)}{2}e^{-a_0p}
\end{equation}
At long times, the cosine within the integral  oscillates very rapidly yielding a vanishing contribution. The remaining integral can be evaluated using the Cauchy theorem resulting in an infinite sum over positive odd Matsubara frequencies. The sum can be performed, yielding
\begin{multline}
\mathcal{H}(x,t\to\infty)=  -\frac{m}{4v}\frac{1}{\pi}e^{-\pi x/v\hbar\beta}\left[\Phi(e^{-2\pi x/v\hbar\beta},1,\frac{1}{2}+\frac{\hbar\beta m}{2\pi})-\Phi(e^{-2\pi x/v\hbar\beta},1,\frac{1}{2}-\frac{\hbar\beta m}{2\pi})\right]\\
+\frac{1}{2\hbar\beta v}\cosech\left(\frac{\pi x}{\hbar v\beta}\right)
-\frac{m}{4v}e^{- m x/v}\tan\frac{\hbar\beta m}{2}
\end{multline}
where $\Phi(x,y,z)$ is the Lerch function.\cite{gradshteyn_tables} Likewise, the long distance behavior dominated by the lowest (Matsubara) frequency term:
\begin{equation}
\mathcal{H}(x,t\to\infty)\approx -\frac{1}{v\hbar\beta}\left[\frac{\pi^{2}}{(\hbar\beta m)^{2}-\pi^{2}}\right]e^{-\pi vx/\hbar\beta}-\frac{ m}{4v}e^{- m x/v}\tan\frac{\hbar\beta m}{2}
\end{equation}
At long distances, this function decays exponentially, but two competing length
scales appear: $v/m$ and $v\hbar\beta/\pi$. The largest sets the characteristic length of the decay.

\section{Long-time dynamics and the generalized Gibbs ensemble}\label{sec:generalized}

It was recently pointed out by Rigol and coworkers~\cite{rigol07_generalized_gibbs_hcbosons}
that, at least for certain observables like the momentum distribution or
the density, the asymptotic (long-time)
behavior of an integrable system following a quantum quench
can described by adopting the maximum entropy (also
called `subjective') approach to Statistical Mechanics,
pioneered by Jaynes.~\cite{jaynes57_information_theory_statistical_mechanics_I,jaynes57_information_theory_statistical_mechanics_II}
Within this approach, the equilibrium state of a system is
described by  a density matrix
that extremizes the von-Neumann entropy, $S=-\Tr\rho\ln\rho$,
subject to  all possible constraints provided  by the  integrals of motion of the Hamiltonian of the
system. In the case of an integrable system, if  $\{I_{m}\}$ is a set of certain (but not all of the possible)
independent integrals of motion of the system,  this procedure leads to a `generalized' Gibbs ensemble,
described by the following density matrix:
\begin{equation}
\rho_\text{gG} = \frac{1}{Z_\text{gG}} e^{-\sum_m \lambda_m
I_m},\label{eq:density_operator_gG}
\end{equation}
where  $Z_\text{gG} = \Tr e^{-\sum_m \lambda_m I_m}$. The values of the Lagrange
multipliers,  $\lambda_m,$ must be determined from the condition that
\begin{equation}
\langle I_m \rangle_\text{gG} =  {\rm Tr} \left[ \rho_{0} I_m  \right]
=  \langle I_m \rangle.
\label{eq:inic_cond_T0}
\end{equation}
where $\rho_{0}$
describes the initial state of the system, and $\langle \cdots \rangle_\text{gG}$ stands for  the average
taken over the generalized Gibbs ensemble,  Eq.
(\ref{eq:density_operator_gG}). Although  $\rho_i = |\Phi(t = 0) \rangle \langle \Phi(t= 0)|$
in the case of a pure state, as was first used in Ref.~\onlinecite{rigol07_generalized_gibbs_hcbosons},
nothing prevent us from taking $\rho_{0}$ to be an arbitrary mixed state  and in particular
a thermal state characterized by an absolute temperature, $T$.  In this case,  the
Lagrange multipliers will  depend on
$T$ or any other parameter that defines the initial state.

Rigol and coworkers numerically tested   the above
conjecture by studying the quench dynamics of a 1D lattice gas of hard-core bosons
(see Ref.~\onlinecite{rigol07_generalized_gibbs_hcbosons} for
more details). One of us showed analytically~\cite{cazalilla06_quench_LL} that correlations of
Luttinger model also relax to averages taken over this ensemble.
This result for the Luttinger model was extended to include finite temperature fluctuations in the initial
state in Ref. ~\onlinecite{iucci09_quench_LL}.
The question that naturally arises then  is whether the
family of integrable models studied in this work (see Eqs. (\ref{eq:genham}), and
their fermionic equivalences  of Eq. (\ref{eq:H_fermionic}) and
(\ref{eq:H_fermionic_matrix})) relax in agreement with the
mentioned conjecture. In other words, does the average
$\langle O\rangle(t)$  at long times relax to the value
$\langle O\rangle_\text{gG} = {\rm Tr}\:  \rho_\text{gG} \, O$, for any
of the correlation functions considered previously? In what follows we shall
address this question by analyzing quantum quenches in the sGM at zero
temperature. The generalization at finite temperatures should be straightforward,
as discussed in Ref.~\onlinecite{iucci09_quench_LL}

\subsection{Quench from the gapped to the gapless phase in the Luther-Emery limit}

In this case, the type evolution of the system is performed by $H_{0}$ (cf. Eq.~\ref{eq:h0le}),
which is diagonal in the operators  $n_{\alpha}(p) = \, : \psi^{\dag}_{\alpha}(p) \psi_{\alpha}(p):$ ($\alpha = r,l$).
Thus, the generalized Gibbs ensemble is defined by the following set of integrals of motion
$I_m \to I_{\alpha}(p) = n_{\alpha}(p)$. We see immediately that the fact that this
ensemble is diagonal in $n^F_{\alpha}(p)$ means that  the order parameter
$\langle e^{-2i\varphi(x)} \rangle_{\text{gG}} = \langle \psi_{r}(x) \psi_{l}(x) \rangle_{\text{gG}} = 0$,
which agrees with the $t \to +\infty$ limit of the order parameter, was shown in Sect~\ref{sect:LE}
to exhibit an exponential decay to zero. However, the two-point
correlator of  $e^{2i \varphi(x)}$
has a non-vanishing limit for $t \to +\infty$. Thus,
our main concern here will be the calculation of the correlation function:
\begin{align}
\langle e^{-2i\varphi(x)}e^{2i\varphi(0)}\rangle_{\text{gG}}
&= \langle \psi^{\dag}_r(x)\psi_l(x) \psi^{\dag}_l(0) \psi_r(0) \rangle_{\text{gG}}\\
&= \sum_{p_{1,}p_{2},p_{3},p_{4}}\frac{ e^{i(p_{1}-p_{2})x}}{L^{2}} \langle
\psi_{r}^{\dagger}(p_{1})\psi_{l}(p_{2})\psi_{l}^{\dagger}(p_{3})\psi_{r}(p_{4})\rangle_{\text{gG}}
\label{eq:twopointle}
\end{align}
Since the ensemble is diagonal in the chirality index, $\alpha$,
as well as momentum, $p$, we evaluation of the above expression
can be carried out by noting that:
\begin{equation}
\langle \psi_{\alpha}^{\dagger}(p)\psi_{\alpha}(p')\rangle_{\text{gG}}
=\frac{\Tr e^{-\sum_{p,\alpha'}\lambda_{\alpha'}(p)
I_{\alpha'}(p)}\psi_{\alpha}^{\dagger}(p)\psi_{\alpha}(p)}
{\Tr e^{-\sum_{p,\alpha'}\lambda_{\alpha'}(p)I_{\alpha'}(p)}}
=  \frac{\delta_{p,p'}}{e^{\lambda_{\alpha}(p)}+1},
\end{equation}
where the Lagrange multipliers $\lambda(q)$
can be related to the values of the same expectation values in the
initial states by imposing their conservation, that is,
\begin{align}
\langle \psi_{l}^{\dagger}(p)\psi_{l}(p)\rangle_{\text{gG}} & =
\frac{1}{e^{\lambda_{l}(p)}+1} = \langle \psi_{l}^{\dagger}(p)\psi_{l}(p) \rangle
=\cos^{2}\theta({p})\\
\langle \psi_{r}^{\dagger}(p)\psi_{r}(p) \rangle_{\text{gG}}  &  =\frac{1}{e^{\lambda_{r}(p)}+1} = \langle \psi_{r}^{\dagger}(p)\psi_{r}(p)\rangle =\sin^{2}\theta({p}).
\end{align}
Hence,
\begin{align}
\langle \psi_{r}^{\dagger}(p_{1})\psi_{l}(p_{2})\psi_{l}^{\dagger}(p_{3})\psi_{r}(p_{4})\rangle_{\text{gG}}  &=\langle\psi_{r}^{\dagger}(p_{1})\psi_{r}(p_{4})\rangle _{\text{gG}}
\times\langle\psi_{l}(p_{2})\psi_{l}^{\dagger}(p_{3})\rangle_{\text{gG}}  \\
&=\delta_{p_{1},p_{4}} \delta_{p_{2},p_{3}}\sin^{2}\theta(p_{1})\left(1-\cos^{2} \theta(p_{2})\right)  \\
&=\delta_{p_{1},p_{4}}\delta_{p_{2},p_{3}}\sin^{2}\theta(p_{1}) \sin^{2}\theta(p_{2}).
\end{align}
Introducing the last expression into Eq.~(\ref{eq:twopointle}) yields:
\begin{equation}
\langle e^{-2i\varphi(x)}e^{2i\varphi(0)}\rangle_{\text{gG}} =
\left\vert \frac{1}{L} \sum_{p} e^{ipx}  \sin^2 \theta({p})\right\vert^2
\end{equation}
and using that $\sin^2\theta(p) = (1 - \cos^2 2 \theta(p))/2$ and
$\cos 2\theta(p) = \omega_0(p)/\sqrt{\omega^2_0(p)+m^2}$, we find
(for $x \neq 0$),
\begin{equation}
\langle e^{-2i\varphi(x)}e^{2i\varphi(0)}\rangle_{\text{gG}} =  \left(\frac{m}{2\pi v}\right)^2
\left[K_1\left(\frac{m |x|}{v}\right)\right]^2,
\end{equation}
which is equal to the $t\to +\infty$ limit of Eq.~(\ref{eq:twopointle3}).
\subsection{Quench from the gapless to the gapped phase in the Luther-Emery limit}
In this case  the initial state is the
gapless ground state of $H_0$, Eq.~(\ref{eq:h0le}), whereas the
Hamiltonian that performs the time evolution has a gap in the spectrum
and it is diagonal in the basis of the $\psi_v(p)$ and $\psi_c(p)$ Fermi operators
(cf. Eq.~\ref{hgapped}).  Therefore, the conserved quantities are
\begin{align}
I_{v}(p)  & = n_v(p) = \psi_{v}^{\dagger}(p)\psi_{v}(p), \\
I_{c}(p)  & = n_c(p) = \psi_{c}^{\dagger}(p)\psi_{c}(p)
\end{align}
The associated Lagrange multipliers (at zero temperature), $\lambda_v(p)$ and $\lambda_c(p)$
can be obtained upon equating $\langle I_{v,c}(p) \rangle_\text{gG} = \langle \Psi(0)| I_{v,c}(p)
| \Psi(0) \rangle$. This yields:
\begin{align}
\langle I_{v}(p)\rangle_\text{gG} &=\frac{1}{e^{\lambda_{v}(p)}+1} =\vartheta(-p)\sin^{2}\theta({p})+\vartheta(p)\cos^{2}\theta({p}) \label{eq:equality_gibbs_1}\\
\langle I_{c}(p) \rangle_\text{gG}&=\frac{1}{e^{\lambda_{c}(p)}+1}
= \vartheta(-p)\cos^{2}\theta({p})+\vartheta(p)\sin^{2}\theta({p})\,,
\label{eq:equality_gibbs_2}
\end{align}
where $\vartheta(p)$ denotes the step function.
Using these expressions we next proceed to compute the
expectation values of the following observables:
\subsubsection{Order parameter}

We start by computing the order parameter,
\begin{equation}
\langle e^{-2i\varphi(x)}\rangle_\text{gG}  =\langle\psi
_{R}^{\dagger}(x)\psi_{L}(x)\rangle
=\frac{1}{2L}\sum_{p}\sin2\theta({p})\left[   \langle
I_{c}(p)\rangle_\text{gG}  -\langle I_{v}
(p) \rangle_\text{gG}\right],
\end{equation}
and upon using Eqs. (\ref{eq:equality_gibbs_1}) and
(\ref{eq:equality_gibbs_2}),

\begin{equation}
\langle e^{-2i\varphi(x)}\rangle_\text{gG}
=-\frac{1}{L}\sum_{p>0}\sin2\theta({p})\cos2\theta({p})\label{eq:le_opq1}
= -\int_{0}^{\infty}\frac{dp}{2\pi}\,
\frac{m \omega_{0}(p)}{\omega^2(p)} e^{-p a_0}
= A(m a_0)
\end{equation}
where we have used that $\cos2\theta_{-p}=-\cos2\theta_{p}$;
$A(m a_0)$ is the non-universal constant introduced
in Sect.~\ref{sec:gaplesstogapped}. This result agrees with
the one obtained in Sect.~\ref{sec:gaplesstogapped} for
the order parameter in the limit $t \to +\infty$.

\subsubsection{Two-point correlation function}

We next consider the two-point correlator of the order
parameter, namely
\begin{equation}
\langle e^{2i\varphi(x)}e^{2i\varphi(0)}\rangle_\text{gG} =
\frac{1}{L^{2}}\sum_{p_{1,}p_{2},p_{3},p_{4}}e^{i(p_{1}-p_{2})x} \langle\psi
_{r}^{\dagger}(p_{1})\psi_{l}(p_{2})\psi_{l}^{\dagger}(p_{3})\psi_{r}
(p_{4})\rangle_\text{gG}. \label{eq:le_gG2pt}
\end{equation}
The calculation of the  average in this case is a bit more  involved,
but it can be performed by resorting to a factorization akin to
Wick's theorem. This is applicable only in the thermodynamic limit,
as it neglects terms in which the four momenta of the above expectation value are equal. These terms yield contributions of $O(1/L)$ compared
the others.  When factorizing as dictated by Wick's theorem,  the only
non-vanishing terms are:
\begin{multline}
\langle\psi_{r}^{\dagger}(p_{1})\psi_{l}(p_{2})\psi_{l}^{\dagger}
(p_{3})\psi_{r}(p_{4})\rangle_\text{gG} =-\delta_{p_{1}p_{4}}\delta_{p_{2}
p_{3}} \langle \psi_{r}^{\dagger}(p_{1})\psi_{r}(p_{4})\rangle_\text{gG}
\langle\psi_{l}^{\dagger}(p_{3})\psi_{l}(p_{2})\rangle_{gG}\\
+\delta_{p_{1}p_{2}}\delta_{p_{4}p_{3}}\langle\psi_{r}^{\dagger
}(p_{1})\psi_{l}(p_{2})\rangle_\text{gG} \langle\psi_{l}^{\dagger}
(p_{3})\psi_{r}(p_{4})\rangle_\text{gG}.
\end{multline}
Upon using
\begin{align}
\langle\psi_{r}^{\dagger}(p)\psi_{r}(p)\rangle_\text{gG}  &=\frac{1}{2}\vartheta(p)\sin^{2}2\theta({p})+\vartheta(-p)\left(  1-\frac{1}{2} \sin^{2}2\theta({p})\right),\\
\langle\psi_{l}^{\dagger}(p)\psi_{l}(p)\rangle_\text{gG} &=\frac{1}{2}\vartheta(-p)\sin^{2}2\theta({p})+\vartheta(p)\left(  1-\frac{1}{2} \sin^{2}2\theta({p})\right), \\
\langle\psi_{l}^{\dagger}(p)\psi_{r}(p)\rangle_\text{gG}&=\langle\psi_{r}^{\dagger}(p)\psi_{l}(p)\rangle_\text{gG}
=-\frac{1}{2}\sin2\theta({p})\cos2\theta({p})\sign (p),
\end{align}
the average over the generalized Gibbs ensemble
of the four Fermi fields on the right hand-side of
Eq.~(\ref{eq:le_gG2pt}) can be computed and yields
the following expression for the two-point correlation function
(up to terms of $O(1/L^2)$):
\begin{equation}
\langle e^{2i\varphi(x)}e^{-2i\varphi(0)}\rangle_\text{gG} = \left\vert -\frac{1}{L}\sum_{p>0}\sin2\theta(p) \cos2\theta(p)\right\vert^2
+\left\vert\frac{1}{L}\sum_{p}e^{ipx}\left[  \frac{1}{2}\vartheta(p)
\sin^{2}2\theta({p})  +  \vartheta(-p)\left(  1-\frac{1}{2}\sin^{2}2\theta_{p}\right)
\right]  \right\vert^2\label{eq:le_q2}
\end{equation}
The first term in r.h.s. of the above expression is just $\langle e^{2i\varphi(x)}\rangle_\text{gG} \:
\langle e^{-2i\varphi(0)}\rangle_\text{gG}$ (cf. Eq.~\ref{eq:le_opq1}),
whereas the first term in the right hand-side
can be written as
\begin{align}
\left\vert \frac{1}{L}\sum_{p}e^{ipx}\left[  \theta(-p)+\frac{1}{2}\sign (
p)\sin^{2}2\theta_{p} \right]\right\vert^2 &
= \left\vert \frac{1}{L} \sum_{p > 0} e^{-ipx} +  \frac{i}{L}\sum_{p > 0} \sin p x  \: \frac{m^2}{\omega^2(p)} \right\vert^2\\
&=  \left\vert \frac{1}{L} \sum_{p > 0} e^{-ipx} + \lim_{t \to +\infty} {\cal H}(x,t) \right\vert^2,
\end{align}
that is, it coincides with the $t \to +\infty$ limit of the second term in the
right hand-side of  Eq.~\ref{eq:two_points_LE}
in Sect.~\ref{sec:le_gaplesstogapped} (the function $\mathcal{H}(x,t)$
is defined in Eq.~\ref{eq:calh}).

\subsection{Quench from the gapless to the gapped phase in the semi-classical limit}

In this case  Hamiltonian  performing the time evolution is gapless ($H_0$)
and thus diagonal in the $b$-operators. Hence, the conserved quantities are
\begin{equation}
I(q)=b^{\dagger}(q)b(q)
\end{equation}
The  Lagrange multipliers of the corresponding generalized
Gibbs density matrix are fixed from the condition:
\begin{align}
\langle I(q)\rangle_\text{gG} &=\frac{1}{e^{\lambda(q)}-1}
= \langle \Phi(0)| b^{\dagger}(q)b(q) | \Phi(0)\rangle\\
&=\sinh^{2}\beta(q), \label{eq:lagrange_sc}
\end{align}
where $\beta(q)$ is defined by Eq.~(\ref{eq:betaqq})
Hence, using this result we next proceed to compute the
order parameter and the two-point correlation function.
We first note that the order parameter vanishes in the generalized Gibbs
ensemble since $\langle e^{-2 i \phi(x)} \rangle_\text{gG} =
e^{- 2 i \langle  \phi^2(0) \rangle_\text{gG} }$ and  $\langle \phi^2(0) \rangle_\text{gG}
=  \frac{\kappa^2}{4} \langle \varphi^2(0) \rangle$ is divergent in the
$L \to +\infty$ limit (see below).
This agrees with the  result found in Sect.~\ref{sec:sc_gappedtogapless}, where it was found
that the order parameter decays exponentially in time.  Thus, in what follows
we shall be concerned with the the two-point correlation function.

\subsubsection{Two-point correlation function}

Since $\langle e^{-2 i\phi(x)} e^{2 i \phi(0)} \rangle_\text{gG} =  e^{-\frac{\kappa^2}{2}
{\cal C}^\text{gG}(x)}$, where $C^{gG}(x) = \langle \varphi(x) \varphi(0) \rangle_\text{gG}
- \langle \varphi^2(0) \rangle_\text{gG}$.  In order to obtain this correlator, we introduce
the Fourier expansion of $\varphi(x)$ (ignoring the zero-mode part),
\begin{equation}
\varphi(x)= \frac{1}{2} \sum_{q\neq0}\left(  \frac{2\pi v}{\omega_0(q) L}\right)
^{1/2}e^{iqx}\left[  b(q)+b^{\dag}(-q)\right],
\end{equation}
into the expectation value, and using (\ref{eq:lagrange_sc})  to evaluate the
averages in the generalized Gibbs ensemble,
we find that, in the thermodynamic limit,
\begin{equation}
\langle \varphi(x) \varphi(0) \rangle_\text{gG} = \int_{0}^{\infty}\frac{d(vq)}{4 \omega_0(q)}  \: \cos qx \: \cosh 2 \beta(q),
\end{equation}
and therefore,
\begin{align}
{\cal C}^{gG}(x) &= \langle\varphi(x)\varphi(0)\rangle_\text{gG} - \langle \varphi^2(0) \rangle_\text{gG} = -  \int_{0}^{\infty}\frac{d ( v q)}{\omega_{0}(q)} \: \cosh 2\beta(q) \: (1-\cos  q x )  \\
&= {\cal C}(x,0) -  \frac{m^2}{4} \int^{+\infty}_0 \frac{d(vq)}{\omega(q) \left[ \omega_0(q)\right]^2} (1-\cos q x)
\end{align}
where ${\cal C}(x,0) \equiv {\cal C}(x,t=0)$ is defined in Eq.~(\ref{eq:cx0}).
Upon comparing the last result  with Eq.~(\ref{eq:cc_mslmsv})
in the limit where $t \to +\infty$, we see they are identical.

\subsection{Quench from the gapless to a gapped phase}

In this case   the Hamiltonian that performs the time evolution is gapped, whereas
the initial state is gapless.  Thus, differently from the previous case, the Hamiltonian
that performs the evolution is diagonal in the $a$-operators, and therefore, the
conserved  quantities are $I(q)=a^{\dagger}(q)a(q)$. The corresponding Lagrange
(at zero temperature) are fixed from the condition:
\begin{align}
\langle I(q)\rangle_\text{gG}  & =\frac
{1}{e^{\lambda(q)}-1}=\left\langle \Phi(0) | a^{\dag}(q) a(q)| \Phi(0) \right\rangle\\
& =\sinh^{2}\beta(q),
\end{align}
where $\beta(q)$ is given by Eq.~(\ref{eq:betaqq}).

In order to obtain the one and  two-point correlation functions
of  $e^{2 i \phi(x)} = e^{2 i \kappa \varphi(x)}$
we first need to write the $\varphi(x)$ field in terms of the $a$-operators. Upon
using the canonical transformation Eq.~(\ref{eq:bogol}):
\begin{equation}
\varphi(x) = \frac{1}{2}\sum_{q\neq 0} \left(  \frac{2\pi}{\omega(q) L} \right)^{1/2} e^{i qx} \left[a(q) + a^{\dag}(-q) \right].
\end{equation}
Hence, since $ \langle e^{-2i \phi(x)} \rangle_\text{gG} =
\langle e^{- i \kappa \varphi(x) } \rangle_\text{gG} = e^{ - \frac{\kappa^2}{2} \langle \varphi^2(0) \rangle_\text{gG}}$,
and $\langle \varphi^2(0) \rangle_\text{gG}$ is logarithmically divergent in the thermodynamic limit (see expressions below),
the find that  $\langle e^{- i \kappa \varphi(x) } \rangle_\text{gG} = 0$. This result is in agreement with the one found in
Sect.~\ref{sec:gaplesstogapped} for the order parameter.

\subsubsection{Two-point correlation function}

Next we consider the two-point correlation function of the same
operator, namely $\langle e^{-2 i \phi(x)} e^{2i \phi(0)} \rangle_\text{gG} = e^{-\frac{\kappa^2}{2} {\cal C}^\text{gG}(x)}$,
where ${\cal C}^\text{gG}(x) = \langle \varphi(x) \varphi(0) \rangle_\text{gG} - \langle \varphi^2(0)\rangle_\text{gG}$.
We first obtain:
\begin{align}
\langle \varphi(x)\varphi(0)\rangle_\text{gG}
&=  \int_{0}^{\infty}\frac{d(vq)}{4 \omega(q)}  \: \cos qx \: \cosh 2 \beta(q).
\end{align}
Hence,
\begin{align}
{\cal C}^\text{gG}(x)  &=   - \int_{0}^{\infty}\frac{ d (vq)}{2\omega(q)}\:  \cosh \beta(q)   \left( 1- \cos qx \right)\: \\
 &= {\cal C}(x,0) + \frac{m^2}{4}\int^{+\infty}_{0} \frac{d(v q)}{\omega_0(q) [\omega(q)]^2} (1 - \cos q x)
\end{align}
where ${\cal C}(x,0)$ is defined in Eq.~(\ref{eq:cc_msl_msv_inic}).
The latter result agrees with Eq.~(\ref{eq:ctc0}) in the $t \to +\infty$ limit.

\section{Relevance to experiments}\label{sec:exp}

As we mentioned in the introduction, ultracold atomic systems
are the ideal arena to study the quench dynamics of isolated quantum many-body systems. This is because
they can be treated, to a large extent,  as entirely isolated systems. Furthermore, since this work is concerned
with the quench dynamics of a specific one dimensional model, the  quantum sine-Gordon model, it is
also worth emphasizing that the properties of these systems are highly tunable and, in particular, so is
their effective dimensionality.  Thus, there are
already a number of experimental realizations of one-dimensional (1D) systems (see \emph{e.g.} Refs.~
\onlinecite{koehl04_1D_bose_gases_opt_lat,paredes04_tonks_gas,kinoshita04_1D_tonks_gas_observation}  and references therein),
and in particular, there are also experiments where non-equilibrium dynamics has been probed in one-dimension,
\emph{e.g.} Refs.~\onlinecite{stoeferle04_shaking_fast_tunnability,kinoshita06_non_thermalization}
Thus, there may be a good chance that some of the results obtained above may be relevant to current or
future quench experiments with ultracold atoms.  However,  since the sine-Gordon model considered
in previous sections is nothing but an effective (low-energy) description of certain  1D systems,
any comparison must be  done with great care, as there is no fundamental reason why the low-energy
effective theory should capture the essentials of the (highly non-equilibrium) quench dynamics. This is
to be contrasted with the equilibrium dynamics, where renormalization group arguments show that
the sine-Gordon model is indeed sufficient to describe the (universal) properties of certain
1D physical systems. There is in fact much evidence, both analytical and numerical, accumulated over the years, of the latter
fact. However, we are not in a comparable situation in the case of non-equilibrium dynamics, and thus
future studies should try to address this question more carefully.

With the above caveat, let us proceed to mention a few situations where the
sGM is applicable,  at least as a good description of the equilibrium state of a
system that can be realized with ultracold atomic systems. There are basically two
kinds of systems, depending on the interpretation of the order parameter
operator, $e^{-2i\phi(x)}$. The fist instance is a 1D Bose gas moving in a
periodic potential, where the  the sine-Gordon model is the
effective field-theory description  the Mott insulator to superfluid transition (MI to SF) in 1D~\cite{haldane81_effective_harmonic_fluid_approach,giamarchi04_book_1d}.
In this case, the order parameter field is the periodic component of the
boson density. A quantum quench  from the gapped (gapless) into the gapless  (gapped) in this system can be realized by suddenly turning on
(off) the periodic potential applied to the 1D gas.  The evolution of the 1D density could be monitored by
performing in-situ measurements, and the two-point correlations by measuring, at different times, the (instantaneous)
structure factor using  Bragg spectroscopy.

 In the second instance, the  order parameter field is interpreted as the (relative) phase of two~\cite{donohue01_bosonic_ladders} (or more\cite{ho04_deconfinement,cazalilla06_deconfinement_longpaper}) 1D Bose gases coupled by Josephson coupling of two 1D Bose gases. Thus, in this setup a quench experiment~\cite{gritsev07_spectroscopy_quench}
from the gapped (gapless)  into the gapless (gapped) phase would
correspond to suddenly switching on (off) the (Josephson)   tunneling, which can be achieved by controlling the
(optical or magnetic trapping) potentials that confine the atoms to 1D. This should be done with care, ensuring
the atoms remain in the 1D regime both in the initial and final states, that is, that \emph{e.g.} the
potential trapping the atoms transversally is always sufficiently tight. The evolution of the relative phase
can be  monitored by analyzing the interference fringes at different times.


\section{Conclusions}\label{sec:conclusions}

To sum up, we have investigated the time evolution of two-point correlation functions and the order parameter after a quantum quench of the relevant operator term in the sine-Gordon model. We considered two different kinds of quenches: a quench from a gapped phase to a gapless phase and viceversa. In addition to the initial pure state, we considered an initial mixed state coupled to an energy reservoir at finite temperature, that is suddenly disconnected at the same time that the quench is performed. In order to compute correlation functions, we studied two limits in which the Hamiltonian renders quadratic in terms of either fermion or boson operators, and the dynamics can be solved exactly: the Luther-Emery  and the semiclassical limits. In the quench from the gaped to the gapless phase, the order parameter decays exponentially for long times to the ground state value, with a time scale fixed by the gap. In turns, the correlation function exhibit a light-cone effect: it decays exponentially with time with a time-scale fixed by the gap until it relaxes to a value that decreases exponentially with distance with correlation length fixed also by the mass. These results are valid for both, the Luther-Emery and the semiclassical limits, and agree with the results obtained in Ref.~\onlinecite{calabrese06_quench_CFT,calabrese07_quench_CFT_long} . For an initial state at high temperature, these decays are also exponential (and the light-cone effect is preserved), but the characteristic time and length is set by the temperature. In between, a crossover connects these two limiting cases.

A general statement for the long time dynamics when the term that opens the gap is suddenly turned on is elusive, since in the semiclassical approximation the long times limit of the correlation function exhibit power-law decay with distance, and the order parameter vanishes, being these features characteristic of a critical state. On the other hand, the results in the Luther-Emery limit indicate a relaxation to a non-universal constant value for long times, signaling an ordered state. The latter behavior at the Luther-Emery limit is also present at finite temperature. Whether these differences are due to an artifact introduced by the semiclassical approximation or to a very special behavior that occurs at the solvable point needs further clarification.

We have shown that the long time behavior of correlation functions and the order parameter in the different types of quenches can be obtained from the generalized-Gibbs ensemble~\cite{rigol07_generalized_gibbs_hcbosons} in which the conservation of a certain set of independent integrals of motion is fixed as a constrain for the maximization of the statistical entropy. Finally, the relevance of the quantum quench dynamics in the sine-Gordon model for cold atomic gases is discussed. The superfluid-Mott insulator transition appears as the most appropriate scenario to observe the described effects.

\acknowledgments

We thank T. Giamarchi and A. Muramatsu for useful discussions. AI gratefully acknowledges financial support from CONICET and UNLP and hospitality of DIPC, where part of this work was done. MAC thanks M. Ueda for his kind hospitality at the University of Tokyo during his visit at the Ueda  ERATO Macroscopic Quantum Control Project of JST (Japan),  during which parts of this manuscript were completed. MAC also gratefully acknowledges financial support of the  Spanish  MEC through grant No. FIS2007-66711-C02-02 and CSIC through grant No. PIE 200760/007.

\appendix

\section{Some identities involving the Bessel functions}

In this appendix we prove the identities\begin{align}
\sum_{n\in\mathbb{Z}}e^{in\phi}K_{0}\left(\alpha\sqrt{\mu^{2}+n^{2}}\right) & =\pi\sum_{l\in\mathbb{Z}}\frac{\exp\left[-\mu\sqrt{\alpha^{2}+(2\pi l+\phi)^{2}}\right]}{\sqrt{\alpha^{2}+(2\pi l+\phi)^{2}}}\label{eq:bessel_identity_1}\\
\sum_{n\in\mathbb{Z}}e^{in\phi}\frac{\alpha}{\sqrt{\mu^{2}+n^{2}}}K_{1}\left(\alpha\sqrt{\mu^{2}+n^{2}}\right) & =\frac{\pi}{\mu}\sum_{l\in\mathbb{Z}}\exp\left[-\mu\sqrt{\alpha^{2}+(2\pi l+\phi)^{2}}\right]\label{eq:bessel_identity_2}\end{align}
where $\phi\in[0,2\pi)$. Using the standard integral representation\begin{equation}
K_{0}(\alpha z)=\frac{1}{2}\int_{-\infty}^{\infty}dx\frac{\cos x\alpha}{x^{2}+z^{2}},\end{equation}
we have\begin{align}
\sum_{n\in\mathbb{Z}}e^{im\phi}K_{0}\left(\alpha\sqrt{\mu^{2}+n^{2}}\right) & =\frac{1}{2}\sum_{n\in\mathbb{Z}}e^{in\phi}\int dk_{x}\frac{e^{ik_{x}\alpha}}{\sqrt{k_{x}^{2}+\mu^{2}+n^{2}}}\\
& =\frac{1}{2\pi}\sum_{n\in\mathbb{Z}}\int dk_{x}\int dk_{y}\frac{e^{ik_{x}\alpha+in\phi}}{k_{x}^{2}+k_{y}^{2}+n^{2}+\mu^{2}}\\
& =\frac{1}{2\pi}\sum_{n\in\mathbb{Z}}\int dk_{x}\int dk_{y}\int dk_{z}\delta(k_{z}-n)\frac{e^{ik_{x}\alpha+ik_{z}\phi}}{k_{x}^{2}+k_{y}^{2}+k_{z}^{2}+\mu^{2}}\end{align}
Next we employ the Poisson summation technique in the form $\sum_{n\in\mathbb{Z}}\delta(k_{z}-n)=\sum_{l\in\mathbb{Z}}e^{2\pi ilk_{z}}$,
and thus\begin{align}
\sum_{n\in\mathbb{Z}}e^{im\phi}K_{0}\left(\alpha\sqrt{\mu^{2}+n^{2}}\right) & =\frac{1}{2\pi}\sum_{l\in\mathbb{Z}}\int dk_{x}\int dk_{y}\int dk_{z}\frac{e^{ik_{x}\alpha+ik_{z}\phi+2\pi ilk_{z}}}{k_{x}^{2}+k_{y}^{2}+k_{z}^{2}+\mu^{2}}\\
& =\frac{1}{2\pi}\sum_{l\in\mathbb{Z}}\int d^{3}k\frac{e^{i\mathbf{k}\cdot\mathbf{r}_{l}}}{k^{2}+\mu^{2}}\end{align}
where we introduced the notation\begin{align}
r_{x} & =\alpha,\\
r_{y} & =0,\\
r_{z} & =\phi+2\pi l.\end{align}
Using a standard expression for the three-dimensional bosonic propagator, we finally obtain
\begin{equation}
\sum_{n\in\mathbb{Z}}e^{im\phi}K_{0}\left(\alpha\sqrt{\mu^{2}+n^{2}}\right)=\pi\sum_{l\in\mathbb{Z}}\frac{e^{-\mu r_{l}}}{r_{l}}.
\end{equation}
where $r_l=|\mathbf{r}_l|$. From this equation the first identity Eq. (\ref{eq:bessel_identity_1}) immediately follows.

To prove the second identity Eq. (\ref{eq:bessel_identity_2}) we proceed by integration on both sides of Eq. (\ref{eq:bessel_identity_1}):
\begin{align}
\int d\alpha\alpha\sum_{n\in\mathbb{Z}}e^{in\phi}K_{0}\left(\alpha\sqrt{\mu^{2}+n^{2}}\right) & =-\sum_{n\in\mathbb{Z}}e^{in\phi}\frac{\alpha}{\sqrt{\mu^{2}+n^{2}}}K_{1}\left(\alpha\sqrt{\mu^{2}+n^{2}}\right)\label{eq:int_K1_1}\\
& =\int d\alpha\alpha\pi\sum_{l\in\mathbb{Z}}\frac{\exp\left[-\mu\sqrt{\alpha^{2}+(2\pi l+\phi)^{2}}\right]}{\sqrt{\alpha^{2}+(2\pi l+\phi)^{2}}}\\
& =-\frac{\pi}{\mu}\sum_{l\in\mathbb{Z}}\exp\left[-\mu\sqrt{\alpha^{2}+(2\pi l+\phi)^{2}}\right]+C'(\phi)\label{eq:int_K1_2}
\end{align}
Thus, from Eqs. (\ref{eq:int_K1_1}) and (\ref{eq:int_K1_2}),
\begin{equation}\label{eq:identity2_C}
\sum_{n\in\mathbb{Z}}e^{in\phi}\frac{\alpha}{\sqrt{\mu^{2}+n^{2}}}K_{1}\left(\alpha\sqrt{\mu^{2}+n^{2}}\right)=\frac{\pi}{\mu}\sum_{l\in\mathbb{Z}}\exp\left[-\mu\sqrt{\alpha^{2}+(2\pi l+\phi)^{2}}\right]+C(\phi)
\end{equation}
$C(\phi)$ is determined from the behavior of $K_{1}(z)$ for $z\to0$:
\begin{equation}
K_{1}(z)\sim\frac{1}{z}.
\end{equation}
and thus, in the limit $\alpha\to0$, on one hand,
\begin{equation}
\lim_{\alpha\to0}\sum_{n\in\mathbb{Z}}e^{in\phi}\frac{\alpha}{\sqrt{\mu^{2}+n^{2}}}K_{1} \left(\alpha\sqrt{\mu^{2}+n^{2}}\right) =\sum_{n\in\mathbb{Z}}\frac{e^{in\phi}}{\mu^{2}+n^{2}}=\frac{\pi}{\mu}\frac{\cosh\mu(\pi-\phi)}{\sinh\mu\pi}
\end{equation}
on the other,
\begin{equation}
\lim_{\alpha\to0}\frac{\pi}{\mu}\sum_{l\in\mathbb{Z}}\exp\left[-\mu\sqrt{\alpha^{2}+(2\pi l+\phi)^{2}}\right] =\frac{\pi}{\mu}\sum_{l\in\mathbb{Z}}\exp\left[-\mu\left|2\pi l+\phi\right|\right] =\frac{\pi}{\mu}\frac{\cosh\mu(\pi-\phi)}{\sinh\mu\pi}
\end{equation}
Therefore, both limits coincide and $C(\phi)=0$, and Eq. (\ref{eq:identity2_C}) reduces to the second identity.


\end{document}